\documentclass[11pt]{article}

\usepackage{cite}
\usepackage{graphicx}
\usepackage{setspace}
\usepackage{amssymb}
\usepackage{amsmath}
\usepackage{textcomp}
\usepackage{fancyhdr}
\usepackage{verbatim}
\usepackage{enumerate}
\usepackage{cancel}
\usepackage{subfig}
\usepackage{anysize}
\usepackage{array}
\usepackage{rotating}
\usepackage{geometry}
\usepackage{feynmp}
\usepackage{authblk}
\usepackage{slashed}
\usepackage{xcolor} 
\usepackage[colorlinks,citecolor=blue]{hyperref}

\def\DO{D$\slashed{0}$}
\def\sss{\scriptscriptstyle}
\def\sR{{\sss R}}
\def\sL{{\sss L}}

\title{Top quark forward-backward asymmetry in $R$-parity violating 
supersymmetry}
\author{Grace Dupuis\thanks{dupuisg@physics.mcgill.ca}\ } 
\author{James M. Cline\thanks{jcline@physics.mcgill.ca}}
\affil{\textit{Department of Physics, McGill University, 3600 
Rue University, Montr\'{e}al, Qu\'{e}bec, Canada, H3A 2T8}}
\date{} 

\begin{document}

\maketitle

\begin{abstract} The interaction of bottom squark-mediated top quark pair
production, occuring in the $R$-parity violating minimal supersymmetric standard
model (MSSM), is proposed as an explanation of the anomalously large $t\bar t$
forward-backward asymmetry (FBA) observed at the Tevatron.   
We find that this model can
give a good fit to top quark data, both the inclusive and invariant mass-dependent
asymmetries, while remaining consistent (at the 2-$\sigma$ level) with the total and differential production
cross-sections.  The scenario is challenged by strong constraints from atomic parity
violation (APV), but we point out an extra diagram for the effective down quark-$Z$
vertex, involving the same coupling constant as required for the FBA, which tends to
weaken the APV constraint, and which can nullify it for reasonable values of the top
squark masses and mixing angle. Large contributions to flavor-changing neutral currents
can be avoided if only the third generation of sparticles is light.

 \end{abstract}

\section{Introduction}
\label{intro}

Recent analyses by both the CDF and \DO\  collaborations have reported an anomalously
large forward-backward asymmetry (FBA) in top pair production at the Tevatron. The
observed discrepancy between the Tevatron data and the standard model (SM) prediction
indicates the possibility of new physics. Within the SM, the top
quark FBA is predicted to be identically zero at leading order (LO), with a nonzero
asymmetry being generated at next-to-leading order (NLO), from extra gluon radiation
and interference of one-loop box diagrams with the LO gluon exchange process. This
predicted NLO asymmetry however, is much lower than that which is observed at
Tevatron.

Reports from the CDF collaboration give inclusive asymmetries, measured in the
$t\bar{t}$ centre of mass frame, of $A^{t\bar{t}}_{FB}=0.155 \pm 0.048$ \cite{Moriond},
$A^{t\bar{t}}_{FB}=0.158 \pm 0.074$ \cite{Aaltonen:2011kc}, $A^{t\bar{t}}_{FB}=0.24 \pm
0.14$ \cite{Aaltonen:2008hc} and $A^{t\bar{t}}_{FB}=0.42 \pm 0.15 \pm 0.05$
\cite{CDFnote10436}, with the last value corresponding to a semileptonic decay channel,
while \DO\  has reported a value of $A^{t\bar{t}}_{FB}=0.196 \pm 0.060
^{+0.018}_{-0.026}$ \cite{Abazov:2011rq, D0ConfProc}. The results display a significant
discrepancy with the NLO SM prediction of
$A^{SM}_{FB}=0.0724^{+0.0104+0.0020}_{-0.0067-0.0027}$ \cite{Ahrens:2011uf}. Moreover,
CDF observed a rise in the asymmetry with both the invariant top mass and top rapidity,
reporting the following values \cite{Moriond}
\begin{equation*}
\begin{split}
A_{\ell} \equiv A^{t\bar{t}}_{FB}(M_{t\bar{t}} < 450 \; \text{GeV})&=0.078 \pm 0.054  \\
A_{h} \equiv A^{t\bar{t}}_{FB}(M_{t\bar{t}} \ge 450 \; \text{GeV})&=0.296 \pm 0.067 
\end{split}
\end{equation*}
compared with the NLO SM predictions of \cite{Ahrens:2011uf} 
\begin{equation*}
\begin{split}
A^{t\bar{t}}_{SM}(M_{t\bar{t}} < 450 \; \text{GeV})&= 0.052^{+0.009}_{-0.006} \\
A^{t\bar{t}}_{SM}(M_{t\bar{t}} \ge 450 \; \text{GeV})&= 0.111 ^{+0.017}_{-0.009}  
\end{split}
\end{equation*}
The \DO\  collaboration does not observe this same rise in the asymmetry with
invariant mass, although the same excess in the inclusive asymmetry is
observed\cite{Abazov:2011rq}. 

Various models of new physics have been proposed to account for the FBA.  They 
are typically classified by the nature of the new mediator particles, 
depending upon whether they are 
$t$- or $s$-channel exchange of vector or scalar bosons. In
general, $s$-channel models involve exchange of a colour-octet vector,
either 
in chiral colour models proposing axigluon-mediated top pair production \cite{
Antunano:2007da, Sehgal:1987wi, Bagger:1987fz, Ferrario:2009bz, Frampton:2009rk,
Chivukula:2010fk, Bai:2011ed, Zerwekh:2011wf, Haisch:2011up, Tavares:2011zg,
Alvarez:2011hi, AguilarSaavedra:2011ci }, or Randall-Sundrum models and the
associated effects of Kaluza-Klein gluons in warped extra dimensions \cite{
Djouadi:2009nb, Bauer:2010iq, Delaunay:2011vv, Park:2009cs}, as well as general
extensions and exotic representations of the SM colour gauge group
\cite{Chen:2010hm, Alvarez:2010js, Barreto:2011au, Foot:2011xu}. 
These models have the advantage of a
large interference with SM gluon exchange. Additionally, a colour-octet only
requires nonzero axial couplings to produce the FBA, whereas 
colour-singlets require both vector and axial-vector couplings to $q \bar{q}$ and $t
\bar{t}$ \cite{Kamenik:2011wt}.

$t$-channel models involve the exchange of scalar or vector mediators whose
couplings can either be flavor-conserving or have large flavor violation. In the
latter case, the generation of a sufficient asymmetry requires a large $u$-$t$ or
$d$-$t$ coupling and concerns then arise regarding the generation of large
flavor-changing neutral currents (FCNCs) \cite{Kamenik:2011wt}. 
Flavor-conserving models
\cite{Ligeti:2011vt, Grinstein:2011yv, Nelson:2011us, Bauer:2009cc, Arnold:2009ay,
Grinstein:2011dz} are advantageous in this respect.
Flavor-violating models typically include extensions of the SM
electroweak gauge group, involving the exchange of new $W^{\prime}$ or
$Z^{\prime}$ bosons with vector-like \cite{Craig:2011an, Bhattacherjee:2011nr, Jung:2009jz,
Ko:2011vd, Cheung:2009ch, Xiao:2010hm, Cheung:2011qa, Berger:2011ua,
Duraisamy:2011pt, Cao:2011ew, Chen:2011mga, Barger:2011ih} or chiral 
\cite{Frank:2011rb, Barger:2010mw, Shelton:2011hq} couplings.

Any new physics contributing to the FBA is significantly constrained by complementary
top quark data, from both the Tevatron and the LHC, which agree with SM predictions.  
Such data in particular includes the total inclusive $t \bar{t}$
production cross-section, $\sigma_{t\bar{t}}$, and the $t\bar{t}$ invariant mass
distribution of the differential top pair cross-section, $d\sigma /dM_{t\bar{t}}$. 
CDF finds a total production cross-section of $\sigma_{t\bar{t}}=(6.9 \pm 1.0)$ pb at 4.6
fb$^{-1}$ \cite{Aaltonen:2009iz}, which agrees with the NLO and NNLO SM predictions.
Reference \cite{Ahrens:2011mw} gives a NNLO value of
$(\sigma_{t\bar{t}}=6.63^{+0.00+0.33}_{-0.40-0.24})$ pb, obtained using the MSTW2008
parton distribution function set \cite{Martin:2009iq}. 

Many $t$-channel scalar exchange models have been noted to 
produce too great a contribution to $d\sigma/dM_{t\bar{t}}$ at
high invariant masses, if they are tuned to give a large FBA
in this regime \cite{Ligeti:2011vt,Gresham:2011pa,Grinstein:2011dz,Blum:2011fa}.
This contribution is of particular concern when considering the LHC data 
corresponding to this observable, which are more sensitive to the effect 
at high invariant masses.
An additional LHC constraint is provided by measurements of the charge asymmetry, 
a quantity which is directly related to the Tevatron FBA, 
which show agreement
with the SM prediction. A recent ATLAS result gives $A_{c}=-0.019 \pm 0.028 \pm 0.024$
\cite{ATLAS:2012an}, compared with a SM prediction of $A_{c}^{SM}=0.0115 \pm 0.0006$
\cite{Rodrigo:2012as}.
 Moreover other models (including the one featured in this paper, it will be seen) are
challenged by constraints from FCNCs such as $B_d$-$\bar B_d$ mixing
\cite{Chivukula:2010fk}, and electroweak precision data.  

In the present work, we focus on the $\tilde b_R^*\, \bar t_R\, d_R^c$ coupling of
 the $R$-parity violating, minimal supersymmetric standard model (MSSM) as
an explanation of the  top quark FBA.\footnote{As this work was nearing completion,
ref.\ \cite{Allanach:2012tc} appeared,  which did a similar analysis, however
without considering in detail the atomic parity violation constraint.  We became aware
of ref.\ \cite{Moriond} by reading this paper.  At the same time, the study
\cite{Hagiwara:2012gy} appeared, which considers essentially the same model, though without emphasizing its
relation to the MSSM and also without reference to APV.}\ \  While $R$-parity conservation
is a simple way to avoid catastrophic baryon violation in the MSSM such as proton
decay, the $\tilde b_R^*\, \bar t_R\, d_R^c$ operator by itself does not lead to
proton decay.  Its coupling is constrained by neutron-antineutron oscillations
\cite{Chang:1996sw}, but the loop diagram giving rise to this constraint is
suppressed if the Wino mass is large or if the squark mixing is small, as we shall
assume in this work.  Other bounds on $R$-parity interactions are on products of
different couplings and can be evaded if the one of interest is dominant.   Such 
operators also have the  virtue of helping to provide a mechanism for
baryogenesis \cite{Cline:1990bw}.

Although the $R$-parity violating  model falls in the category of those 
claimed to be disfavored
because of the tension with the asymmetry at high invariant masses, we
believed it had sufficiently strong theoretical motivation to suspend judgment on
this issue.  In the meantime, the more recent measurement of $A_{h}$
reported in \cite{Moriond} greatly ameliorates this tension.   In agreement with
\cite{Allanach:2012tc,Hagiwara:2012gy}, we find that the $\tilde{b}$-mediated down quark-antiquark
annihilation process can account for the observed value of the FBA, while remaining
consistent with the constraints posed by the  total and differential production
cross-sections. 

It was recently pointed out \cite{Gresham:2012wc} that  atomic parity
violation (APV) provides an additional powerful low-energy  constraint on
$t$-channel models of the FBA.   The same couplings of $u$-$t$ or $d$-$t$ to  the
new mediator that generate the FBA can generically produce large anomalous couplings
of $u$ and $d$ to $Z$ that contribute to the weak charge of nuclei. We agree that 
for generic values of the stop mass matrix, the
APV constraints from cesium rule out the regions of parameter space
compatible with the FBA in the $R$-parity violating MSSM; however, if there is 
significant stop mixing and a large mass splitting, we show that there exists a
 canceling contribution
to the anomalous parity violation which can greatly relax the constraints and preserve
the viability of the model for explaining the top quark observations.

In section \ref{tprod}, we describe the computation of the FBA and  in section
\ref{apv} the new contributions to APV.  Results of our numerical analysis are
presented in  section \ref{res}, and conclusions given in \ref{conc}, including a brief
discussion of the problems of FCNCs and perturbativity of the large $R$-parity violating coupling.

\section{Top pair production in $R$-parity violating SUSY}
\label{tprod}

\begin{figure}
\centering
\begin{fmffile}{SquarkExchange}
\parbox{60mm}{\begin{fmfgraph*}(60,30)
	\fmfleft{i1,i2}
	\fmfright{o1,o2}
	
	\fmflabel{$d$}{i1}
	\fmflabel{$\bar{d}$}{i2}
	\fmflabel{$t$}{o1}
	\fmflabel{$\bar{t}$}{o2}
	
	\fmf{fermion}{i1,v1,i2}
	\fmf{fermion}{o2,v2,o1}
	
	\fmf{gluon}{v1,v2}
\end{fmfgraph*}} + \parbox{60mm}{\begin{fmfgraph*}(60,30)
	\fmfleft{i1,i2}
	\fmfright{o1,o2}
	
	\fmflabel{$d$}{i1}
	\fmflabel{$\bar{d}$}{i2}
	\fmflabel{$t$}{o1}
	\fmflabel{$\bar{t}$}{o2}
	
	\fmf{fermion}{v2,i2}
	\fmf{phantom}{v2,o2}
	\fmf{fermion}{i1,v1}
	\fmf{phantom}{v1,o1}
	\fmf{dots,lab.side=left, label=$\tilde{b}$}{v1,v2}
	
	\fmf{fermion,tension=0}{o2,v1}
	\fmf{fermion,tension=0}{v2,o1}

\end{fmfgraph*}}

\end{fmffile}
\vskip0.5cm
\caption{Feynman diagrams showing both the SM (left) and MSSM (right) contributions to the process 
$d\bar{d} \to t\bar{t}$.}
\label{fig:fig0c}
\end{figure}
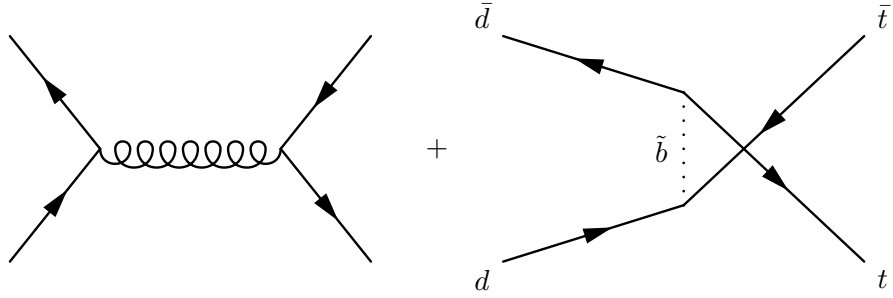

The MSSM without $R$-parity is supplemented by the additional 
superpotential contributions
\begin{equation}
\mathcal{W}_{\cancel{R}}=\frac{1}{2}\lambda_{ijk}L_{i}L_{j}E_{k}^{c}+\lambda'_{ijk}\delta^{\alpha \beta}L_{i}Q_{j\alpha}D_{k\beta}^{c}+\frac{1}{2}\lambda''_{ijk}\epsilon^{\alpha\beta\gamma}U_{i\alpha}^{c}D_{j\beta}^{c}D_{k\gamma}^{c}+\mu_{i}L_{i}H_{2}
\end{equation}
where the chiral superfields
$L_i$ and $Q_i$ represent the left-handed lepton and quark doublets,
$E_i$ and $U_{i},\,D_{i}$ are the right-handed lepton and quark singlets, and
$H_{1,2}$ are the Higgses.\footnote{The superscript $c$ denotes charge conjugation and the 
indices $(i,j,k)$ and $(\alpha,\beta,\gamma)$ correspond respectively to 
generation and colour}\ \    For the top quark FBA however, we are only interested 
in the baryon-violating $\lambda''$ term, which gives
rise to $d\bar{d} \rightarrow t\bar{t}$ via $\tilde b$ exchange, 
through the first of the two interactions
\begin{equation}
   \mathcal{L}_{\lambda''}=-\lambda''_{ijk}\epsilon_{\alpha\beta\gamma} 
   \left [ (\tilde{d}^{k\gamma}_{R})^{\dagger} \bar{u}^{i\alpha}_{R}
   (d^{j\beta}_{R})^{c} + \frac{1}{2}(\tilde{u}^{i\alpha}_{R})^{\dagger} 
   \bar{d}^{j\beta}_{R}(d^{k\gamma}_{R})^{c} \right ] + {\rm h.c.},
   \label{eqn:eqnWR}
\end{equation}
or more specifically (focusing on the relevant generations for the FBA and APV)
\begin{equation}
   \mathcal{L}_{\rm int}=f_{dt}\,\epsilon^{\alpha\beta\gamma}\, \left(
   \tilde{b}^{\dagger}_{\gamma}\, \bar{t}_{\alpha} P_{L} d_{\beta}^{c}
	+ \tilde{t}^{\dagger}_{\gamma}\, \bar{b}_{\alpha} P_{L} d_{\beta}^{c},
	\right)  + {\rm h.c.},
   \label{Lint}
\end{equation}
where $P_{R,L}=(1 \pm \gamma_{5})/2$ denote the chirality projection operators and the
coupling ${-}\lambda''_{313}$ is now denoted by $f_{dt}$.
The corresponding Feynman diagram for the new contribution to $d\bar d\to t\bar t$ is shown in Figure
\ref{fig:fig0c}.   It involves only the first interaction in (\ref{Lint}), while the
second enters into the computation of APV observables to be discussed below.

Structurally similar models for the $t\bar t\,$ FBA have been considered more
generally under the guise of  scalar, diquark-mediated, $u$-channel top pair
production within a framework of generalized exotic colour representations under the
SM gauge group SU(3)$_{c} \times$ SU(2)$_{L} \times$ U(1)$_{Y}$
\cite{Arhrib:2009hu}. Since it can be shown \cite{Manohar:2006ga} that under the
assumption of minimal flavour violation and avoiding FCNCs, a scalar octet is
associated with quark masses and hence, the corresponding $s$-channel process
mediated by a colour-octet is negligible \cite{Arhrib:2009hu}, the cases of interest
are the colour-triplet and sextet. A particular instance of the colour-triplet,
specifically the representation $(3,1, -1/3)$, has the same quantum numbers as the
bottom squark, and so corresponds to the case of interest in this work.

The interaction (\ref{Lint}) involves only the right-handed squarks. In general,
one expects mixing between the left and right squarks due to off-diagonal
terms in the mass matrix.   For a given flavor of squark, these terms are 
proportional to the corresponding quark mass, and thus squark
mixing is generally only considered for the third generation. For our FBA
analysis, we neglect mixing of the bottom squark, but we will allow for 
top squark mixing when we discuss a loophole to the atomic parity violation
constraints.

Detailed formulas for our computation of the FBA are given in appendix
\ref{fbacomp}.

\section{Atomic parity violation}
\label{apv}

\begin{figure}[b]
\centering
\begin{fmffile}{APV2}
\parbox{60mm}{\begin{fmfgraph*}(60,30)
	
	\fmfleft{i1,i2}
	\fmfright{o1}
	
	\fmflabel{$d$}{i1}
	\fmflabel{$\bar{d}$}{i2}
	\fmflabel{Z}{o1}
	
	\fmf{fermion}{i1,v1}
	\fmf{fermion}{v3,i2}
	\fmf{fermion}{v3,v2}
	\fmf{fermion,label=$t,,b$}{v2,v1}
	\fmf{photon}{v2,o1}
	
	\fmf{dots, left=0.5, tension=0.2, lab.side=left, label=$\tilde{b},,\tilde{t}$}{v1,v3}
	
\end{fmfgraph*}} $\qquad+$ \parbox{60mm}{\begin{fmfgraph*}(60,30)
	
	\fmfleft{i1,i2}
	\fmfright{o1}
	
	\fmflabel{$d$}{i1}
	\fmflabel{$\bar{d}$}{i2}
	\fmflabel{Z}{o1}
	
	\fmf{fermion}{i1,v1}
	\fmf{fermion}{v3,i2}
	\fmf{dots,lab.side=left,label=$\tilde{b},,\tilde{t}$}{v3,v2}
	\fmf{dots,lab.side=left,label=$\tilde{b},,\tilde{t}$}{v2,v1}
	\fmf{photon}{v2,o1}
	
	\fmf{fermion, left=0.5, tension=0.2, lab.side=left, label=$t,,b$}{v1,v3}
	
\end{fmfgraph*}}

\end{fmffile}
\vskip0.5cm
\caption{Anomalous coupling of the $Z$ boson to $d$ quark at one loop, 
arising from $\tilde b^* \bar d^c_R t_R$ and 
$\tilde t^* \bar b^c_R t_R$ couplings.  Left: standard contribution,
present with no squark mixing;
right: extra contributions arising from nonzero squark mixing.  Note that the
diagrams with $b$ and $\tilde t$ in the loop arise from the second term in
(\ref{Lint}).}
\label{fig:fig0d}
\end{figure}
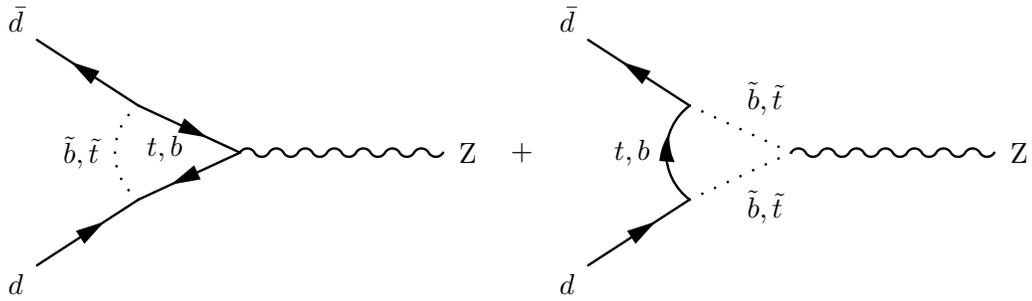

As mentioned in the introduction, the $\tilde b^* \bar d^c_R t_R$ coupling
also affects the one-loop effective $\bar d_\sR \slashed{Z} d$
vertex and hence is subject to the constraints of atomic parity
violating data. The relevant Feynman diagrams are
 shown in Figure \ref{fig:fig0d}.   In the case of vanishing squark mixing, only 
the first diagram contributes to the anomalous coupling of
the right-handed down quark to the $Z$
\cite{Gresham:2012wc},\footnote{Ref.\ \cite{Gresham:2012wc} computed $a_{R,u}$
for a closely-related model with coupling to $u_\sR$ rather than $d_\sR$
(the case referred to as mediator $M=\omega$ in that reference), but the 
result is the same in the present model}
\begin{equation}
  a_{R,d}^{NP}=\frac{f_{dt}^{2}}{16\pi^{2}} \left( F_1(m_{t}^{2}/m_{\tilde{b}}^{2})
  - F_1(m_{b}^{2}/m_{\tilde{t}}^{2}) \right),
  \label{anp}
\end{equation}
where $F_1$ is defined as
\begin{equation}
   F_1(x) \equiv x\frac{(x-1-\ln{x})}{(1-x)^{2}}.
   \end{equation}
The coefficient is normalized such that the corresponding one-loop effective 
Lagrangian is
\begin{equation}
	{\cal L}_{\rm eff} = -{g_2\over \cos\theta_W}  
a_{R,d}^{NP}\, Z^\mu\, \bar d_R\gamma_\mu d_R
\label{Leff}
\end{equation}
where $g_2$ is the SU(2)$_L$ coupling and $\theta_W$ is the Weinberg angle.
Since $F_1\sim -x\ln x$ as $x\to 0$, the second term with $x=m_b/m_{\tilde t}$
is negligible.

In the $R$-parity violating MSSM, if there is squark mixing then 
there are necessarily additional contributions from the right-most diagrams
of fig.\ \ref{fig:fig0d}.  The one with $t$ and $\tilde b$ in the loop has the
same sign as the dominant contribution to (\ref{anp}).\footnote{One might have
expected the opposite sign due to opposite weak isospins of $t_\sL$ and
$\tilde b_\sL$, but this is reversed by the factor of $\slashed{p}p_\mu\to
p^2\gamma_\mu/4$ in the
second diagram, which changes sign under Wick rotation.}  Since we are interested
in finding a canceling contribution, let us suppose that the $b$ squark mixing angle
is small, and focus on the diagram with $b$ and $\tilde t$ in the loop.  This
has the opposite sign to (\ref{anp}) as desired.  
We define $\delta$ such that
the squared stop masses are given by $m^2_{\tilde t_{1,2}} = m_0^2(1\pm \delta/2)$.
In terms of the $\tilde t$ mixing angle $\theta_{\tilde t}$, 
 we find that the extra contribution to $a_R^{NP}(d)$ is given by
\begin{equation}
	\delta a_{R,d}^{NP} = -\frac{f_{dt}^{2}}{32\pi^{2}}\sin^2(2\theta_{\tilde t})
	F_2(\delta)
\end{equation}
in the limit where $m_b \ll m_0$, where 
\begin{equation}
	F_2(\delta) = {1\over \delta}\ln\left({2+\delta\over 2-\delta}\right)-1
\end{equation}
Depending upon the properties of the top squark, 
this extra contribution can be significant compared to the first
one, $a_{R,d}^{NP}$.  Suppose for example that the top squark mass splitting is 
large, $\delta = 1.8$, corresponding to masses $m_{\tilde t_{1,2}} = (0.32,1.38)m_0$,
for which $F_2 = 0.64$.  A large cancellation against the first diagram can be
achieved at $m_{\tilde b} = 600$ GeV (relevant for the benchmark model we designate
below), where $2F_1 = 0.31$, if
$\sin^2 2\theta_{\tilde t} \cong 0.5$, which is large but not maximal mixing.  The weakening of
the Cs APV constraint for more general values of $m_{\tilde b}$ will be considered
in the next section.

By $Z$ exchange,
the operator (\ref{Leff}) induces an anomalous contribution to the 4-quark effective operator 
\begin{equation}
    \delta\mathcal{L}=\frac{G_{F}}{\sqrt{2}}\, \delta C_{1d} \,(\bar{e}\gamma^{\mu}
	\gamma_{5}e)\,(\bar{d}\gamma_{\mu}d) 
\end{equation}
where $\delta C_{1d}=(a_{R,d}^{NP}+\delta a_{R,d}^{NP})$, to be compared to its
SM value $C_{1d}=\frac{1}{2}-\frac{2}{3}\sin^2\theta_{W}$.  There is an analogous
term with $C_{1u}=-\frac{1}{2}+\frac{4}{3}\sin^2\theta_{W}$ 
for the up quark, for which $\delta C_{1u}=0$ in our model.
The weak charge of a nucleus with $Z$ protons and $N$ neutrons 
is given by $Q_{W}(Z,N)= -2 [(2Z+N)C_{1u}+(2N+Z)C_{1d}]$ \cite{Bouchiat:1997mj}.
A strong constraint on new physics contributions to APV comes from cesium
($^{133}$Cs), for which the weak charge is measured to be $Q_{W}(Cs)=-73.20(35)$
\cite{PDG,Porsev:2009pr,Porsev:2010de}, compared to the SM prediction of $Q_{W}^{SM}(Cs)=-73.15(2)$
\cite{PDG}.  Improved constraints are anticipated
from the Qweak experiment \cite{Carlini:2010zz}
on the proton nuclear
weak charge, measured via electron-proton scattering.
Current data give $Q_{W}(p)=0.052(17)$, consistent with the expected value of
$Q_{W}^{SM}(p)=0.0713(8)$ \cite{Young:2007zs}.

Further constraints on anomalous parity violation arise from 
neutrino deep inelastic scattering (DIS) experiments. In a similar manner
to the nuclear weak charge case, the contribution of right-handed down quarks
is described by the 4-fermion interaction
\begin{equation}
   \mathcal{L}_{\nu d}= \frac{G_{F}}{\sqrt{2}}\,(\epsilon_{\sR,d}+\delta\epsilon_{\sR,d} )\,
   \,(\bar{\nu}\gamma^{\mu}(1-\gamma_{5})\nu)
   \,(\bar{d}\gamma_{\mu}(1+\gamma_{5})d)
\end{equation}
where the SM contribution is $\epsilon_{\sR,d}= 
(2/3)\sin^{2}\theta_{W}$ and 
 the new physics contribution is
$\delta\epsilon_{\sR,d}= a^{NP}_{R,d}+ \delta a^{NP}_{R,d}$. Experiments in neutrino deep inelastic
scattering provide measurements of the quantities $g_{\sR,\sL}^{2} \equiv \sum_{q}
\epsilon^{2}_{\sR,\sL,q}$.  Current data give 
$g_{R}^{2}=0.0309(10)$, in agreement with the SM
prediction of $0.03001(2)$ \cite{PDG}.  These do not constrain our model as strongly
as do the atomic cesium data.

\section{Numerical Analysis and Results}
\label{res}

In this section we describe our method of analysis and the resulting best-fit 
regions in the $R$-parity violating MSSM parameter space for fitting the FBA as well as top quark
production data.  Then we show how these regions are constrained by atomic
parity violation.

\subsection{Analysis}

Following ref.\ \cite{Grinstein:2011dz}, we distinguish the new physics (NP) contribution
to the FBA from the SM contribution through the definition 
\begin{equation}
   A_{FB}^{NP+SM}=\frac{\sigma_{F}^{NP}-\sigma_{B}^{NP}}{\sigma_{F}^{NP}+\sigma_{B}^{NP}+\sigma_{LO}^{SM}}+
   A^{SM}_{FB} \left ( \frac{\sigma^{SM}}{\sigma^{SM}+\sigma^{NP}}\right),
\label{eqn:ASM}
\end{equation}
where the subscripts $F,B$ denote the forward and backward contributions to the
cross section, respectively, and interference of the NP contribution with the
LO SM term is taken into account in $\sigma_{F,B}^{NP}$, while the
SM NLO$+$NNLO contributions are included in $\sigma^{SM}$ and 
$A^{SM}_{FB}$.

In addition to the inclusive asymmetry, we also consider the FBA coming from 
the low and high invariant mass regions.  From ref.\ \cite{Ahrens:2011uf}, the
 NLO + NNLL values for the SM contributions are 
$A^{SM}_{FB}=0.0724^{+0.0106}_{-0.00722}$, $ A^{SM}_{h}=0.111 ^{+0.017}_{-0.009}$
and $A^{SM}_{\ell}=0.052^{+0.009}_{-0.006}$ respectively, and the NNLO value of the
total SM cross-section is $\sigma^{SM}=(6.63^{+0.33}_{ -0.36})$pb
\cite{Ahrens:2011mw}. The central values of the cross-section for the high and low
invariant mass bins are taken to be $\sigma^{SM}_h = 2.34$ pb and 
$\sigma^{SM}_l = 4.29$ pb. An overall
$K$-factor of 1.4 was used to estimate higher-order QCD contributions
to the cross-sections. $\sigma^{SM}$ includes the contribution of
$gg$ fusion in addition to
the dominant $q\bar q$ annihilation process. To maintain
consistency with references \cite{Ahrens:2011uf} and \cite{Ahrens:2011mw}, from
which the SM predictions are taken, the MSTW2008NLO set of PDFs is used
\cite{Martin:2009iq}, evaluated at a scale $q=m_{t}=173.1$ GeV. The PDF scale is
varied between $m_{t}/2$ and $2m_{t}$ to give an approximate uncertainty on the
calculated values. The top mass is taken to be $m_t = 172.5$ GeV for the 
remainder of the analysis \cite{Aaltonen:2011em}.

To compare the predictions to experimental values,  the inclusive asymmetry is
taken to be $A^{t\bar{t}}_{FB}=0.178 \pm 0.037$, obtained from a weighted
average of the four values quoted in section 1, adding the uncertainties in
quadrature. The values of the folded, invariant mass dependent asymmetries are
taken to be $A_{h}=0.296 \pm 0.067$ and $A_{\ell}=0.078 \pm 0.054$ \cite{Moriond}.
The total production cross-section, $\sigma_{t\bar{t}}=(6.9 \pm 1.0)$pb and
differential cross-section invariant mass distribution are as in ref.\ 
\cite{Aaltonen:2009iz}; in particular, the value for the high-mass bin is 
$(d\sigma/dM_{t\bar{t}})_{h}\equiv d\sigma/dM_{t\bar{t}}(800$ GeV  $\le
M_{t\bar{t}} \le 1400 $GeV$)=(0.068 \pm 0.032 \pm 0.015 \pm 0.004)$ fb GeV$^{-1}$.

We determine the allowed regions in the $(m_{\tilde b_R},f_{dt})$ parameter space
(recall that squark mixing is neglected), by fitting the anomalous experimental
values of the inclusive asymmetry and the corresponding low and high invariant mass
bins, while maintaining agreement with the total and differential production
cross-sections. Following \cite{Bhattacherjee:2011nr} we construct a 
$\chi^{2}$ function in terms of these parameters,
 according to the naive definition
\begin{equation}
\chi^{2}(m_{\tilde{b}}, f_{dt})=\sum_{i} \frac{(\mathcal{O}_{i}(m_{\tilde{b}},f_{dt})-\mathcal{O}_{i}^{exp})^2}{\sigma_{i}^{2}}
\label{eqn:chi}
\end{equation}
where $\mathcal{O}_{i}$ correspond to the Tevatron observables 
(we treat APV and LHC data separately below) and $\sigma_{i}$ their
corresponding uncertainties, which are derived from the uncertainties on the
experimental values and SM predictions, and those estimated from variation of the
PDFs, by adding in quadrature.   This gives a rough estimate of the allowed regions, since the true
$\chi^{2}$ function should take into account the correlations between
these observables through the covariance matrix; 
clearly the cross-section and the various asymmetries are
correlated.   However this is a much more involved task, and for this work we
content ourselves with the estimate provided by (\ref{eqn:chi}).

An additional, separate $\chi^2$ is defined to fit LHC data, specifically the charge asymmetry
in $t\bar{t}$ production and the relative differential top pair production cross-section, $1/\sigma$
$d\sigma/dM_{t\bar{t}}$, at high invariant masses. 
Although there is no forward-backward asymmetry at the LHC due to the
fact that both beams are protons, the charge asymmetry is closely
related, and is defined as 
\begin{equation}
A_{c}=\frac{\sigma(\Delta |y| \ge 0)-\sigma(\Delta |y| \le 0)}{\sigma(\Delta |y| \ge 0)+\sigma(\Delta |y| \le 0)},
\end{equation}
where $\Delta |y|=|y_{t}|-|y_{\bar{t}}|$. The NP and SM contributions to the charge asymmetry are distinguished
as in (\ref{eqn:ASM}). 
The measured value and SM prediction of the charge
asymmetry are as given in sect. \ref{intro}.

We find that a much stronger constraint is posed by the differential cross-section,  $1/\sigma$
$d\sigma/dM_{t\bar{t}}$.   We consider the relative differential cross-section corresponding to the two
highest invariant mass bins,  $700 \text{ GeV} < M_{t\bar{t}} < 950 \text{ GeV}$ and $950 \text{ GeV} <
M_{t\bar{t}} < 2700 \text{ GeV}$, measured by ATLAS as $(0.24 \pm 0.04)$ TeV$^{-1}$ and $(0.007 \pm 0.003)$
TeV$^{-1}$, respectively \cite{ATLAS:2012hg}.  Recently it was pointed out that electroweak Sudakov
corrections lead to a reduction in the predicted cross section in models where the FBA is explained by new
physics in the $t$-channel \cite{Manohar:2012rs}.  This reference showed that the reduction was
approximately 10\% at high invariant $t\bar t$ masses, in models where the new physics couples to
left-handed quarks.  It would be beyond the scope of the present work to recompute this correction for the
case of interest, for new couplings to right-handed quarks.  We simply assume it is the same and thus we
apply a 10\% downward correction to the predicted cross section.

We further define an analogous $\chi^2$ for the atomic parity violation data,
using the three measurements of the Cs and proton weak charges and the neutrino
DIS determination of $g_{R}$, as discussed in sect.\ \ref{apv}.  The resulting constraints on $(m_{\tilde b_R},f_{dt})$
are dominated by the Cs contribution.

\begin{figure}[bt]
\centering
{ \includegraphics[angle=0,width=15cm]{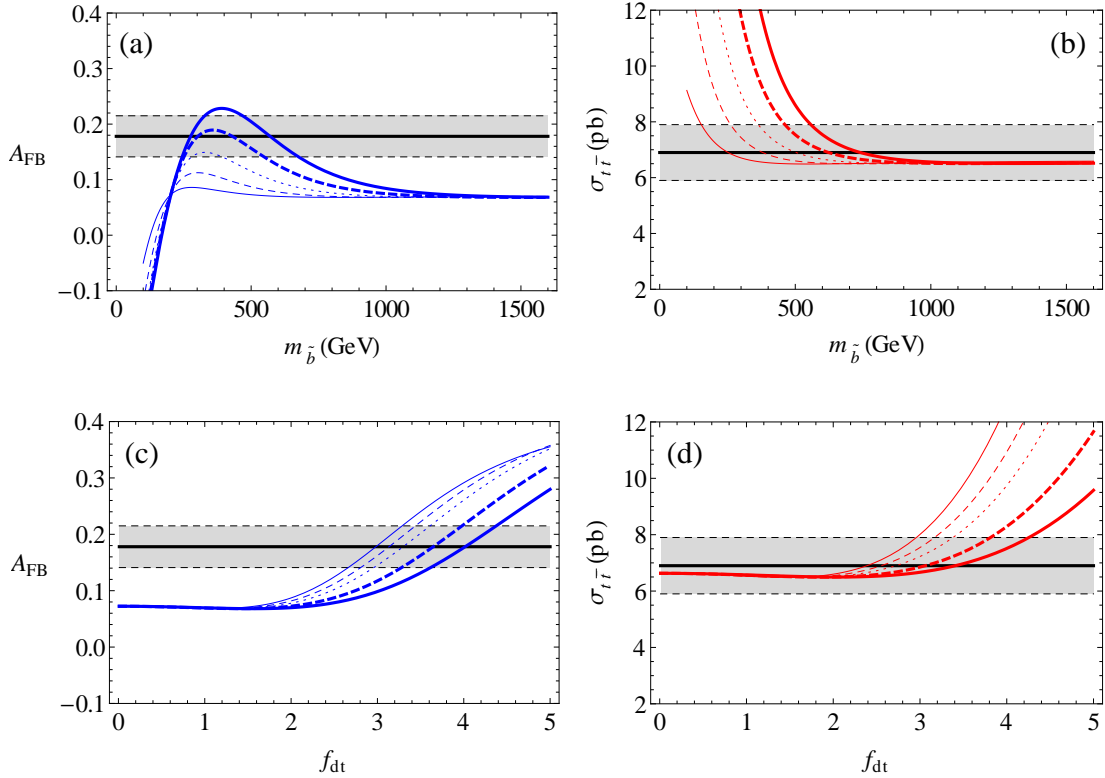}} 
\caption{Figs.\ (a) and (b) show the sbottom mass dependence of the 
$t\bar{t}$ forward-backward asymmetry and $t\bar{t}$ production cross-section 
respectively, for various chosen values of the coupling $f_{dt}$. 
The solid, dashed, dotted, thick-dashed and thick-solid lines ({\it i.e.,} going
from bottom to top) correspond to 
$f_{dt}=\lbrace 1.6, 2.0, 2.4, 2.8, 3.2\rbrace$. The experimental values are 
indicated by the black line, with the grey shaded regions corresponding to 
$1\sigma$ deviation.  Figs.\ (c) and (d) analogously show the dependence upon
$f_{dt}$ for $m_{\tilde{b}}=\lbrace 500, 550,
600, 700, 800\rbrace$ GeV, going from top to bottom.
}
\label{fig:fig1}
\end{figure}

\begin{figure}[tb]
\centering
\includegraphics[angle=0,width=15.0cm]{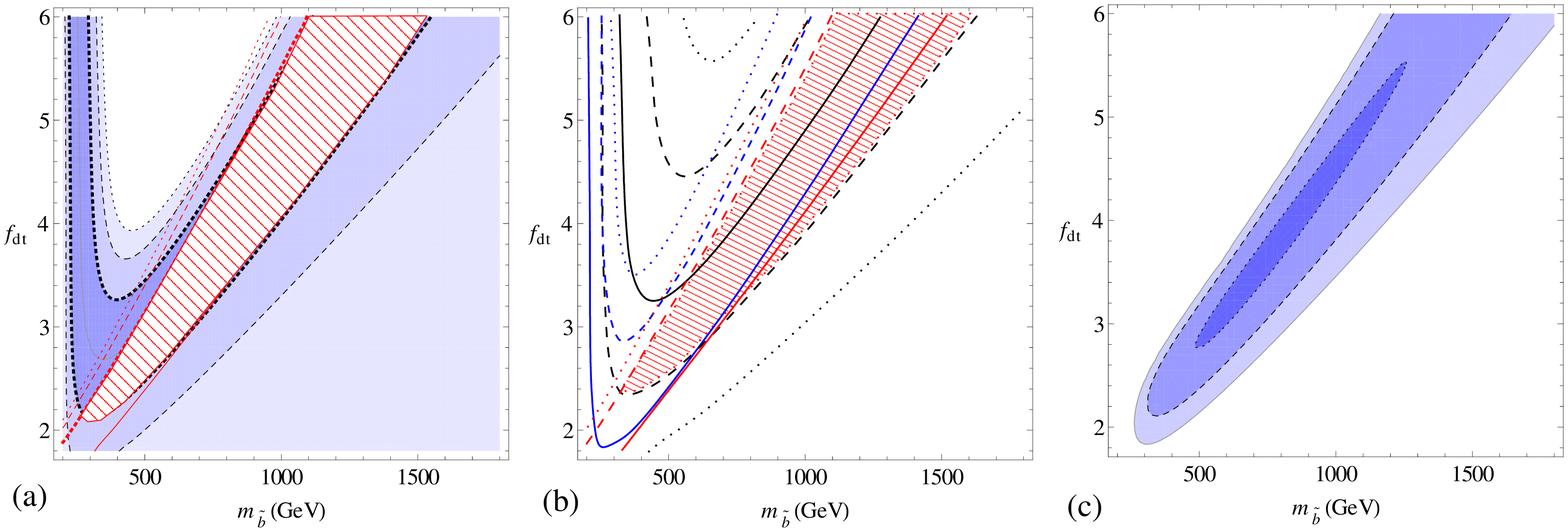}
\caption{(a) Shaded regions are 68\%, 95\% and 99\% confidence intervals in the
$m_{\tilde b}$-$f_{dt}$ plane for
FBA; (red) diagonal lines are central value and 68\%, 95\%, 99\% c.l.\ upper limits on
$f_{dt}$ for the production cross section; region of overlap of the 68\%
intervals is cross-hatched.
(b) Same as (a), but for the low (blue, lower curves) and high (black,
higher curves) invariant mass bins, and omitting the 99\% contour.
Solid lines indicate the central values.  Overlap of all 68\% intervals is cross-hatched. 
(c) Allowed 68\%, 95\% and 99\%  c.l.\ regions from the combined $\chi^2$ for the 
FBA and the production cross section.}
\label{contours}
\end{figure}

\subsection{Results}

We find that the $R$-parity violating MSSM  can generate a sufficient FBA, while
maintaining agreement with the total production cross-section, to within $1\sigma$.
The dependence of both the asymmetry and production cross-section on  $m_{\tilde b}$
is shown in Figure \ref{fig:fig1}(a,b), for various choices of the coupling $f_{dt}$.
Conversely, the dependence of these quantities on the coupling is shown in Figure
\ref{fig:fig1}(c,d) for several  values of $m_{\tilde b}$.  There is a line of
degeneracy in the $f_{dt}$-$m_{\tilde b}$ plane where a simultaneous fit can be
achieved for both quantities, which can be seen in fig.\ \ref{contours}. Here we
plot the allowed intervals for the FBA and the production cross section together,
showing that they overlap in a diagonal region starting near $m_{\tilde b}=400$ GeV,
$f_{dt}=2.4$, and continuing to higher masses and couplings along a band with slope
$df_{dt}/dm_{\tilde b}\cong 0.35/(100$ GeV).   Fig.\ \ref{contours}(a) shows the
separate regions for the FBA and $\sigma_{tt}$ superimposed, while 
\ref{contours}(b) shows the allowed regions for the low and high invariant mass
parts of the FBA.  Fig.\ \ref{contours}(c) displays the overall preferred regions
derived from the $\chi^2$ (\ref{eqn:chi}).  The best-fit value is close to
$m_{\tilde{b}}=600 \mbox{\ GeV}, f_{dt}=3.2$, which we take as our benchmark
model.

\begin{table}[tb]
\begin{tabular}{| c | c | c |}
\hline
& $\mathcal{O}_{i}^{exp}$& $\mathcal{O}_{i}(m_{\tilde{b}}=600 \mbox{ GeV}, f_{dt}=3.2)$ \\
\hline
$A_{FB}^{t\bar{t}}$& 0.178  $\pm$ 0.037& 0.167\\
$A_{\ell}$& 0.078 $\pm$ 0.054& 0.107\\
$A_{h}$& 0.296 $\pm$ 0.067& 0.252\\
$\sigma_{t\bar{t}}$& (6.9 $\pm$ 1.0)pb& 7.54 pb\\
$d\sigma_{t\bar{t}}^{h}/dM_{t\bar{t}}$& (0.068 $\pm$ 0.032 $\pm$ 0.015 $\pm$ 0.004)
fb\,GeV$^{-1}$& 0.097 fb\,GeV$^{-1}$\\
\hline
\end{tabular}
\caption{Values of the fitted observables, for the benchmark model 
$(m_{\tilde b} = 600 \mbox{ GeV},\ f_{dt}=3.2)$ close to the minimum of
the $\chi^2$.}
\label{tab2}
\end{table}

\begin{figure}[tbp]
\centering
\includegraphics[angle=0, width=15.0cm]{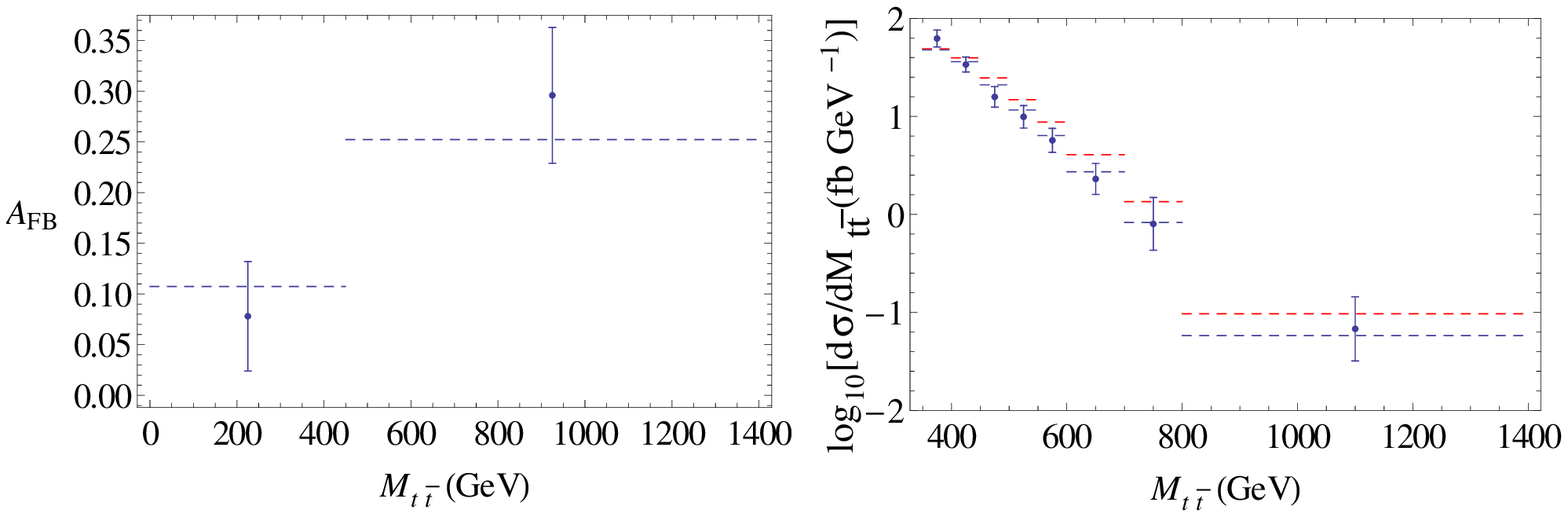}
\caption{(a) Left: the invariant mass-dependent asymmetry; 
the values of the high and low mass bin are displayed both for the CDF result, 
shown with errors, and the calculated value, represented by the dashed line. 
Benchmark values of $m_{\tilde b} = 600 \mbox{ GeV}$, $f_{dt}= 3.2$ were 
chosen, corresponding to the minimum of the total $\chi^{2}$.
(b) Right: binned invariant mass distribution of the differential
$t\bar{t}$ production cross section, displaying both the CDF results (shown with
error bars) and the calculated result, represented by the dashed lines. The 
lower (blue)
dashed line shows the result calculated assuming only the standard model process while
the upper (red) includes the new physics contribution from squark exchange. The same
model parameters are assumed as in (a).}
\label{fig:panel3}
\end{figure}

In addition to the inclusive asymmetry, we also find that the model gives a good
simultaneous fit of the folded asymmetry values, corresponding to the high and low
invariant mass bins, $M_{t\bar{t}} < 450 \mbox{ GeV}$ and $M_{t\bar{t}} \ge 450
\mbox{ GeV}$. This is shown in fig.\  \ref{fig:panel3}(a).  The high and low mass
asymmetries $A_{h}$ and $A_{\ell}$ are within $1\sigma$ of the respective
experimental values. The fit to the full set of collider observables is displayed in
table \ref{tab2}. The goodness of fit is determined, accounting for the total top
pair production cross-section, the differential cross-section at high invariant
masses $(800 \mbox{ GeV} < M_{t\bar{t}} \le 1400 \mbox{ GeV})$, the inclusive
asymmetry, and the asymmetries at high and low invariant masses.   The differential
top pair production cross-section, with respect to the $t\bar{t}$ invariant mass, is
presented in fig.\  \ref{fig:panel3}(b), while simultaneously fitting the full set
of observables considered. It can be seen in the relevant plots that the results
remain consistent with the experimental invariant mass distribution. Moreover, the
problem of the enhancement of the tail of the distribution, which previously plagued
$t$-channel models of this kind 
\cite{Ligeti:2011vt,Gresham:2011pa,Grinstein:2011dz,Blum:2011fa}
is resolved by using the updated value of $A_h$ \cite{Moriond}.
 Both of these figures and the table refer to the
benchmark model $(m_{\tilde{b}}=600 \mbox{\ GeV}, f_{dt}=3.2)$.

We find that there is some tension between this model and LHC data, namely the relative
differential cross-section, as enhancement of the cross-section at high invariant
masses disfavors a significant region of parameter space. In fig. \ref{fig:LHC}
we show the region of parameter space allowed by the LHC-measured observables.
Although the 1-$\sigma$ allowed regions from the two experiments do not intersect, there
is considerable overlap between the 2-$\sigma$ allowed regions, spanning sbottom masses from
300 to 900 GeV and couplings $f_{dt}$ from 2.4 to 3.6.  Thus although the overall fit from
combining Tevatron and LHC is not ideal, the model is far from being ruled out.  We differ 
in this conclusion from ref.\ \cite{AguilarSaavedra:2011vw}, which considered more generally the constraints on models with 
color-triplet exchange.  In that work, an effective operator approach was used, where the heavy mediator
particle was integrated out, which may not be justified
for the low squark masses considered here.  Other differences between the two analyses is our use of
more recent ATLAS data \cite{ATLAS:2012hg} and taking into account electroweak Sudakov corrections
\cite{Manohar:2012rs}.  On the other hand,
the constraint from the charge asymmetry is relatively weak, and we find the model to be consistent to 
within 1$\sigma$ with this measurement over the entire parameter space considered. As the most recent LHC determinations of the charge asymmetry are
dominated by both statistical and systematic uncertainty, it is not yet at the level where it can 
provide significant constraints on models that predict an enhanced FBA.

\begin{figure}[tbp]
\centering
\includegraphics[angle=0, width=14.0cm]{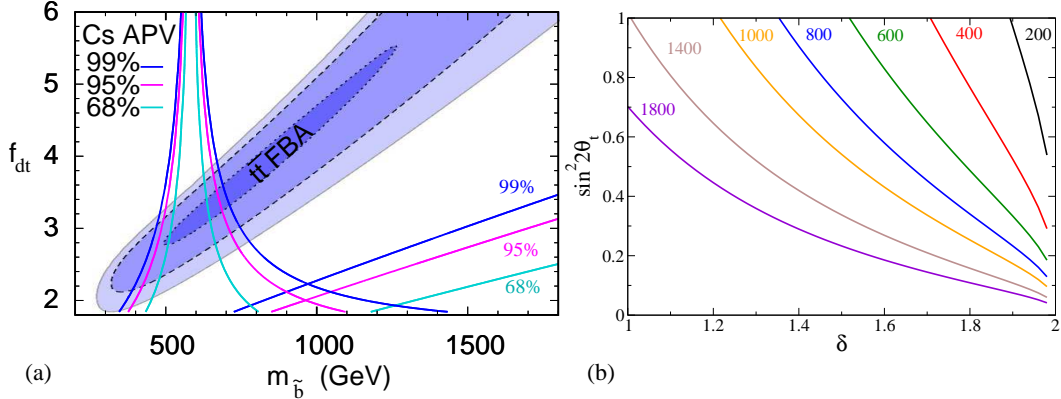}
\caption{(a) Left: upper limits on $f_{dt}$ from Cesium atomic parity violation at 68\%, 95\% and
99\% c.l., superimposed on the (shaded) regions allowed by the top quark FBA and production cross
section.  The lower set of curves (straight lines) assumes a generic squark spectrum, while the
upper set requires large top squark mixing $\sin^2 2\theta=0.5$ and mass splitting
$\delta \equiv (m^2_{\tilde t_1} - m^2_{\tilde
t_2})/[\frac12(m^2_{\tilde t_1} + m^2_{\tilde t_2})] = 1.8$.  (b) Right: relation
between $\sin^2 2\theta_{\tilde t}$ and  $\delta$ needed for nullification
of the Cs constraint, for different values of $m_{\tilde b}$ as labeled on the
figure, in GeV.}
\label{fig:apv}
\end{figure}

\begin{figure}[tbp]
\centering
\includegraphics[angle=0, width=8.0cm]{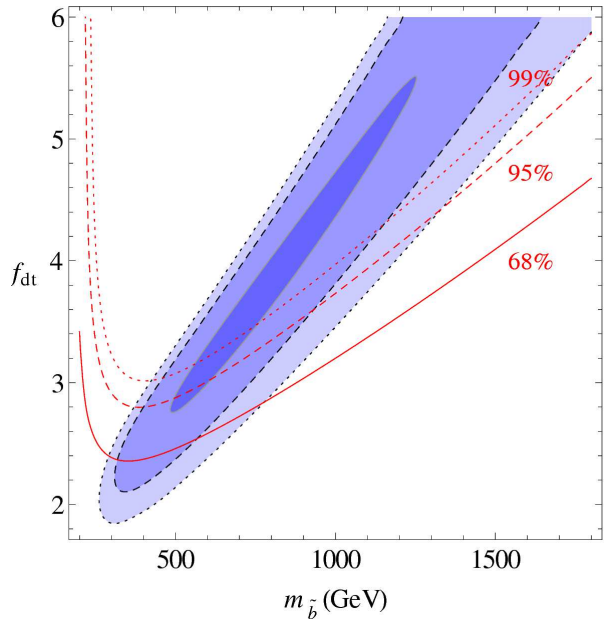}
\caption{(a) Upper limits on the relative differential cross section for $t\bar t$ production
from LHC data at 68\%, 95\% and
99\% c.l., superimposed on the (shaded) regions allowed by the Tevatron
top quark FBA and production cross-section.}
\label{fig:LHC}
\end{figure}

In ref.\ \cite{Gresham:2012wc}, it was shown that the absence of anomalous
parity violation in Cs rules out the explanation of the top quark FBA in some
models that are closely related to the present one.  We confirm that this is also
the case for the $R$-parity violating MSSM, if only the leftmost diagram of 
fig.\ \ref{fig:fig0d} contributes significantly to the $\bar d\sR \slashed{Z} d_\sR$ coupling.
Fig.\ \ref{fig:apv}(a) shows that the upper limits (diagonal lines)
on the coupling $f_{dt}$ as a function
of $m_{\tilde b}$ are well below the values required to explain the FBA in this
case.  However as we noted in section \ref{apv}, the extra diagram with $Z$ coupling
to $\tilde t$ is generically present and contributes with the opposite sign.  
To get a large cancellation with the first diagram
requires both large mixing and large mass splitting of the top
squarks.   In Fig.\ \ref{fig:apv}(a) 
we show an example of the  weakened constraints 
obtained by taking $\sin^2 2\theta_{\tilde t}=0.5$ and stop masses split by
$\delta = 1.8$, corresponding to 
$m_{\tilde t_{1,2}} = (0.32,1.38)m_0$.  (The overall scale $m_0\gg m_b$ is irrelevant for
the value of the rightmost diagram in fig.\ \ref{fig:fig0d}, which is
dimensionless.)  We see in this example that the cancellation is complete for
$m_{\tilde b}\cong 600$ GeV and fully allows for our benchmark model.  More generally,
the value of $m_{\tilde b}$ for which the cancellation is maximized depends upon
the stop mixing and mass splitting, and moves to higher values as we take smaller
splitting.  It is determined by $F_1(m_t^2/m^2_{\tilde b}) = 2\sin^2 2\theta_{\tilde t}
F_2(\delta)$.  This relation is plotted in the plane of $\sin^2 2\theta_{\tilde t}$ and
$\delta$ for the relevant range bottom squark masses in fig.\ \ref{fig:apv}(b).  Thus
one can achieve consistency with the APV constraints over a range of reasonable
values of the parameters.

\section{Conclusion}
\label{conc}

The anomalously large top pair forward-backward asymmetry, observed by both CDF and
\DO\  is suggestive of new physics, although new models are subject to numerous
constraints from collider data, namely the strong agreement between the SM prediction
and experimental result in the total $t\bar{t}$ production cross-section. The
process  $d\bar{d} \to t\bar{t}$, via sbottom exchange in $R$-parity violating
supersymmetry with the interaction $\tilde b_R^*\, \bar t_R\, d_R^c$,
has been proposed as a mechanism to account for the FBA.  We find that
the model can produce the required asymmetry, while preserving agreement with the total
production cross-section. Moreover, results remain consistent with the differential
cross-section, even at  high invariant masses.

\newpage

However, there is some tension between the model and LHC data. Although the model preserves agreement
with the measured value of the charge asymmetry, a large region of the parameter space allowed
by simultaneously fitting the Tevatron observables is disfavored due to enhancement of the relative
differential cross-section at high invariant masses.  Nevertheless there is overlap between 
the 2-$\sigma$ allowed regions of the FBA and the ATLAS top production cross section, leaving
$R$-parity violating SUSY as possible explanation of the anomaly.  The agreement is improved somewhat by
taking into account electroweak Sudakov corrections \cite{Manohar:2012rs}, which we have done in an
approximate way.  A more exact computation of this effect for new physics coupling to right-handed quarks,
as appropriate to the present model, could be interesting for future study.

The constraints posed by atomic parity violation in Cesium would naively appear to 
exclude this explanation, but we have shown that a canceling contribution to the
offending one is naturally present due to the operator 
$f_{dt}\tilde t_R^*\, \bar b_R\, d_R^c$  that automatically comes with the same strength
as $f_{dt}\tilde b_R^*\, \bar t_R\, d_R^c$ in the R-parity violating MSSM, and it can
nullify the the effect of the latter for large but reasonable values of the
stop mixing and mass splitting.  This loophole can also be compatible with a constraint
on $f_{dt}$ from neutron-antineutron oscillations \cite{Chang:1996sw} if the
sbottom quark mixing is small or the Wino mass is large.  It could be interesting to
quantify this statement in light of the allowed region of parameter space we identify
in this work.

An interesting feature of the model could be its implication to $b$ physics, since the
extra operator $f_{dt}\tilde t_R^*\, \bar b_R\, d_R^c$  implies analogous new
contributions to $b\bar{b}$ production, leading to further constraints and possible
predictions for future experimental results.   One expects to observe a
forward-backward asymmetry in $b\bar{b}$ production at a comparable level to that of
$t\bar{t}$.

A possible concern with this model is the existence of large flavor-changing
 neutral currents.\footnote{we realized
this problem as we were finishing the present paper}\ \   For example, there is a box diagram for
$B_d$-$\bar B_d$ mixing with $t$, $\tilde b$ and $\tilde d$ in the loop which induces a
contribution to the effective operator $\Lambda^{-2}(\bar b_\sR\gamma_\mu d_\sR)
(\bar d_\sR\gamma_\mu b_\sR)$ with $\Lambda \cong 8\pi m_{\tilde q}/f_{dt}^2$
assuming a common squark mass $m_{\tilde q}$.  For our benchmark model, this would
give $\Lambda = 1.5$ TeV, which is far below the limit of 200 TeV for models with
no GIM suppression (though close to the limit of 1.8 TeV for models having minimal
or next-to-minimal flavour violating structure) \cite{Bona:2007vi}. However, this outcome can be
avoided if $m_{\tilde d}\ggg m_{\tilde b}$. There are well-motivated SUSY scenarios in which only
the third generation is light \cite{Dimopoulos:1995mi,Cohen:1996vb,Barbieri:2010pd,Craig:2011yk}, 
and it has been argued that $R$-parity violation is
particularly natural within this framework \cite{Brust:2012uf,Brust:2011tb}.
A complete study of FCNC constraints on our model in this case is beyond the scope of this paper,
 but would be interesting to undertake.

After completion of this work, we became aware of ref. \cite{Cao:2009uz}, which also includes a study
of the FBA in the $R$-parity violating MSSM. There the constraint $|f_{dt}|<1.25$ was imposed,
leading to a negative conclusion. This was based upon requiring $f_{dt}$ to remain perturbative
up to the unification scale \cite{Goity:1994dq}. With couplings $f_{dt} \gtrsim 2.8$ as we favor for the FBA,
such a requirement must be abandoned in favor of the more modest one that $f_{dt}$ does not reach
a Landau pole at too low a scale. The paradigm of very heavy first and second generation squarks
may help in this respect, since $f_{dt}$ runs more slowly in their absence, with $df_{dt}/dt \cong 
3f_{dt}^2/16\pi^2$, ignoring gauge couplings. We estimate the Landau pole scale to be 
$\Lambda \sim 14$ TeV for the case $f_{dt}=2.8$, $m_{\tilde{b}}=500$ GeV, going up to 
$\Lambda \sim 70$ TeV for $f_{dt}=2.2$, $m_{\tilde{b}}=300$ GeV (the lower boundary of the
95 $\%$ c.l. region).

\section*{Acknowledgements}
We thank Mike Trott for valuable guidance and suggestions at the beginning of this
project.  We also thank Sean Tulin for very helpful correspondence. 
JC thanks the Aspen Center for Physics where discusssions pertinent to 
this work took place. GD is supported
by Tomlinson and McGill Majors Fellowships.  JC is supported by Natural Sciences
and Engineering Research Council, Canada.

\appendix

\section{Computation of FBA}
\label{fbacomp}
The matrix element for the process $d\bar{d} \to t\bar{t}$, including the SM
contribution by $s$-channel gluon exchange, and neglecting the $\tilde b$ width
(which we reinstate below) is given by
\begin{equation}
\begin{split}
\mathcal{M}_{SM+\tilde{b}}= &\frac{g_{s}^{2}}{\hat{s}}[\bar{v}(p_{\bar{d}})\gamma_{\mu}t^{a}u(p_{d})][\bar{u}(p_{t})\gamma^{\mu}t^{a}v(p_{\bar{t}})] \\
 & - \frac{|f_{dt}|^{2}}{2(\hat{u}-m_{\tilde{b}}^{2})} \biggl \{ [\bar{v}(p_{\bar{d}})\gamma_{\mu}P_{R}u(p_{d})][\bar{u}(p_{t})\gamma^{\mu}P_{R}v(p_{\bar{t}})]  \\
&  -[\bar{v}_{\alpha}(p_{\bar{d}})\gamma_{\mu}P_{R}u^{\beta}(p_{d})][\bar{u}_{\beta}(p_{t})\gamma^{\mu}P_{R}v^{\alpha}(p_{\bar{t}})] \biggr \},
\end{split}
\end{equation}
where $\hat{s}$ and $\hat{u}$ are the Mandelstam variables in the partonic centre of mass frame and the quark momenta are taken as
\begin{equation}
\begin{split}
p_{d,\bar{d}} =& \frac{\sqrt{\hat{s}}}{2} (1,0,0,\pm1) \nonumber \\
p_{t,\bar{t}}=& \frac{\sqrt{\hat{s}}}{2}(1, \pm\beta\sin{\theta}, 0, \pm\beta\cos{\theta}),
\end{split}
\end{equation}
with $\beta^{2}=1-4m_{t}^{2}/\hat{s}$ and where $\theta$ is the scattering angle in the partonic centre of mass frame. Squaring and summing over colour and spin gives
\begin{equation}
\begin{split}
|\mathcal{M}_{SM+\tilde{b}}|^{2}=& 4\pi^{2}\alpha_{s}^{2}\left( \frac{N_{c}^{2}-1}{N_{c}^{2}}\right) \left( 1+\beta^{2}\cos^{2}{\theta}+\frac{4m_{t}^{2}}{\hat{s}}\right) \\
& -\frac{\pi\alpha_{s}}{2} \left( \frac{N_{c}^{2}-1}{N_{c}^{2}}\right) \left( \frac{ |f_{dt}|^{2}\hat{s}}{\hat{u}-m_{\tilde{b}}^{2}}\right) \left(  (1+\beta\cos{\theta})^{2}+\frac{4m_{t}^{2}}{\hat{s}} \right)   \\
 &  + \frac{1}{8} \left( \frac{N_{c}-1}{N_{c}}\right) \left(\frac{ |f_{dt}|^{2}\hat{s}}{\hat{u}-m_{\tilde{b}}^{2}} \right) ^{2} (1+\beta\cos{\theta} )^{2},
\end{split}
\end{equation}
with $N_{c}=3$. Accounting for resonance near the sbottom mass, the matrix element becomes
\begin{equation}
\begin{split}
|\mathcal{M}_{SM+\tilde{b}}|^{2}=& \frac{32\pi^{2}\alpha_{s}^{2}}{9} \left( 1+\beta^{2}\cos^{2}{\theta}+\frac{4m_{t}^{2}}{\hat{s}}\right) \\
& -\frac{4\pi\alpha_{s}|f_{dt}|^{2}}{9}\frac{\hat{s} (\hat{u}-m_{\tilde{b}}^{2}) \left( 1+\beta^{2}\cos^{2}{\theta}+\frac{4m_{t}^{2}}{\hat{s}}\right )}{(\hat{u}-m_{\tilde{b}}^{2})^{2}+\Gamma^{2}m_{\tilde{b}}^{2}} \\
& + \frac{|f_{dt}|^{4}\hat{s}^{2}}{12}\frac{(1+\beta\cos{\theta})^{2}}{(\hat{u}-m_{\tilde{b}}^{2})^{2}+\Gamma^{2}m_{\tilde{b}}^{2}}.
\end{split}
\end{equation}
The sbottom decay width in this model is dominated by $\tilde b\to \bar t \bar d$, giving
$\Gamma \cong |f_{dt}|^{2}(m_{\tilde{b}}/8\pi)$ $\times(1-m_t^2/m_{\tilde b}^2)^2$. The differential cross-section is then given by
\begin{equation}
\frac{d\hat{\sigma}}{d\cos{\theta}}=\frac{\beta}{32\pi s} | \mathcal{M}_{SM+\tilde{b}} |^{2},
\end{equation}
with the hat denoting the partonic quantity in the usual convention.
The forward-backward asymmetry, defined as
\begin{equation}
A_{FB}=\frac{\sigma(\Delta y \ge 0)-\sigma(\Delta y \le 0)}{\sigma(\Delta y \ge 0)+\sigma(\Delta y \le 0)},
\end{equation}
where $\Delta y$ is the rapidity of the top quarks, $y_{t}-y_{\bar{t}}$, is then given in the $t\bar{t}$ centre of mass frame by
\begin{equation}
A_{FB}^{t\bar{t}}=\frac{\sigma_{F}-\sigma_{B}}{\sigma_{F}+\sigma_{B}},
\end{equation}
where $\sigma_{F}$ and $\sigma_{B}$ are the forward and backward hadronic cross-sections, evaluated in the $t\bar{t}$ frame, and given by
\begin{equation}
\sigma_{F}=\int_{0}^{1} d \cos{\theta}\frac{d\sigma}{d\cos{\theta}},  \qquad \sigma_{B}=\int_{-1}^{0} d \cos{\theta}\frac{d\sigma}{d\cos{\theta}},
\end{equation}
with  $\Delta y=2\tanh^{-1}(\beta\cos{\theta})$. The hadronic cross-sections are 
obtained by convolutions with the PDFs, according to
\begin{equation}
\frac{d\sigma}{dM_{t\bar{t}}}=\frac{M_{t\bar{t}}}{s}\int_{-y_{B}^{0}}^{y_{B}^{0}} dy_B \int_{-1}^{1} dz f_d(\tau e^{y_{B}})f_d(\tau e^{-y_{B}}) \frac{d\hat{\sigma}}{dz},
\end{equation}
where $s$ is the hadronic centre of mass energy, $y_{B}$ is the boost rapidity of 
the subprocess frame, $\tau=\sqrt{\frac{M_{t\bar{t}}^{2}}{s}}$, 
$z=\cos{\theta}$ and $f_{i}$ are the parton distribution functions. 
In accordance with \cite{Lane:1991qh}, an experimental rapidity cut 
$|y_{1,2}|< y_{\rm max}$ corresponds to the integration limits
\begin{equation}
y_{B}^{0}={\rm min}[y_{\rm max},-\ln{\tau}].
\end{equation}

\bibliographystyle{utphys}
\bibliography{ttbarFBA}   

\providecommand{\href}[2]{#2}\begingroup\raggedright\begin{thebibliography}{10}

\bibitem{Moriond}
S.~Leone, ``{Top Quark Production at the Tevatron},'' in {\em {Rencontres de
  Moriond, EW}}.
\newblock 2012.
\newblock
  \url{http://indico.in2p3.fr/getFile.py/access?contribId=22&sessionId=9&resId%
=0&materialId=slides&confId=6001}.

\bibitem{Aaltonen:2011kc}
{\bfseries CDF Collaboration} Collaboration, T.~Aaltonen {\em et~al.},
  ``{Evidence for a Mass Dependent Forward-Backward Asymmetry in Top Quark Pair
  Production},'' \href{http://dx.doi.org/10.1103/PhysRevD.83.112003}{{\em
  Phys.Rev.} {\bfseries D83} (2011) 112003},
  \href{http://arxiv.org/abs/1101.0034}{{\ttfamily arXiv:1101.0034 [hep-ex]}}.
23 pages, 18 figures, submitted to Physical Review D.

\bibitem{Aaltonen:2008hc}
{\bfseries CDF Collaboration} Collaboration, T.~Aaltonen {\em et~al.},
  ``{Forward-Backward Asymmetry in Top Quark Production in $p\bar{p}$
  Collisions at $\sqrt{s}=1.96$ TeV},''
  \href{http://dx.doi.org/10.1103/PhysRevLett.101.202001}{{\em Phys.Rev.Lett.}
  {\bfseries 101} (2008) 202001},
\href{http://arxiv.org/abs/0806.2472}{{\ttfamily arXiv:0806.2472 [hep-ex]}}.

\bibitem{CDFnote10436}
{\bfseries CDF} Collaboration, Y.~Takeuchi {\em et~al.}, ``{Measurement of the
  Forward Backward Asymmetry in Top Pair Production in the Dilepton Decay
  Channel using 5.1 fb$^{-1}$},'' 2007.
\newblock \url{http://www-cdf.fnal.gov/physics/new/top/2011/DilAfb/}.

\bibitem{Abazov:2011rq}
{\bfseries D0 Collaboration} Collaboration, V.~M. Abazov {\em et~al.},
  ``{Forward-backward asymmetry in top quark-antiquark production},''
  \href{http://dx.doi.org/10.1103/PhysRevD.84.112005}{{\em Phys.Rev.}
  {\bfseries D84} (2011) 112005},
\href{http://arxiv.org/abs/1107.4995}{{\ttfamily arXiv:1107.4995 [hep-ex]}}.

\bibitem{D0ConfProc}
{\bfseries D0 Collaboration} Collaboration, R.~Demina, ``{Measurement of the
  forward-backward charge asymmetry in top quark production in
  proton-antiproton collisions at 1.96 TeV},'' in {\em {EPS-HEP Conference,
  Grenoble}}.
\newblock 2011.
\newblock
  \url{http://pos.sissa.it/archive/conferences/134/379/EPS-HEP2011_379.pdf}.

\bibitem{Ahrens:2011uf}
V.~Ahrens, A.~Ferroglia, M.~Neubert, B.~D. Pecjak, and L.~L. Yang, ``{The
  top-pair forward-backward asymmetry beyond NLO},''
  \href{http://dx.doi.org/10.1103/PhysRevD.84.074004}{{\em Phys.Rev.}
  {\bfseries D84} (2011) 074004},
\href{http://arxiv.org/abs/1106.6051}{{\ttfamily arXiv:1106.6051 [hep-ph]}}.

\bibitem{Antunano:2007da}
O.~Antunano, J.~H. Kuhn, and G.~Rodrigo, ``{Top quarks, axigluons and charge
  asymmetries at hadron colliders},''
  \href{http://dx.doi.org/10.1103/PhysRevD.77.014003}{{\em Phys.Rev.}
  {\bfseries D77} (2008) 014003},
\href{http://arxiv.org/abs/0709.1652}{{\ttfamily arXiv:0709.1652 [hep-ph]}}.

\bibitem{Sehgal:1987wi}
L.~Sehgal and M.~Wanninger, ``{Forward-Backward Asymmetry in Two Jet Events:
  Signature of Axigluons in Proton-Antiproton Collisions},''
\href{http://dx.doi.org/10.1016/0370-2693(88)91138-0}{{\em Phys.Lett.}
  {\bfseries B200} (1988) 211}.

\bibitem{Bagger:1987fz}
J.~Bagger, C.~Schmidt, and S.~King, ``{AxigluonbProduction in Hadronic
  Collisions},''
\href{http://dx.doi.org/10.1103/PhysRevD.37.1188}{{\em Phys.Rev.} {\bfseries
  D37} (1988) 1188}.

\bibitem{Ferrario:2009bz}
P.~Ferrario and G.~Rodrigo, ``{Constraining heavy colored resonances from
  top-antitop quark events},''
  \href{http://dx.doi.org/10.1103/PhysRevD.80.051701}{{\em Phys.Rev.}
  {\bfseries D80} (2009) 051701},
\href{http://arxiv.org/abs/0906.5541}{{\ttfamily arXiv:0906.5541 [hep-ph]}}.

\bibitem{Frampton:2009rk}
P.~H. Frampton, J.~Shu, and K.~Wang, ``{Axigluon as Possible Explanation for $p
  \bar{p} \rightarrow t \bar{t}$ Forward-Backward Asymmetry},''
  \href{http://dx.doi.org/10.1016/j.physletb.2009.12.043}{{\em Phys.Lett.}
  {\bfseries B683} (2010) 294--297},
  \href{http://arxiv.org/abs/0911.2955}{{\ttfamily arXiv:0911.2955 [hep-ph]}}.
revtex. 5 pages, 2 figures.

\bibitem{Chivukula:2010fk}
R.~S. Chivukula, E.~H. Simmons, and C.-P. Yuan, ``{Axigluons cannot explain the
  observed top quark forward-backward asymmetry},''
  \href{http://dx.doi.org/10.1103/PhysRevD.82.094009}{{\em Phys.Rev.}
  {\bfseries D82} (2010) 094009},
\href{http://arxiv.org/abs/1007.0260}{{\ttfamily arXiv:1007.0260 [hep-ph]}}.

\bibitem{Bai:2011ed}
Y.~Bai, J.~L. Hewett, J.~Kaplan, and T.~G. Rizzo, ``{LHC Predictions from a
  Tevatron Anomaly in the Top Quark Forward-Backward Asymmetry},''
  \href{http://dx.doi.org/10.1007/JHEP03(2011)003}{{\em JHEP} {\bfseries 1103}
  (2011) 003},
\href{http://arxiv.org/abs/1101.5203}{{\ttfamily arXiv:1101.5203 [hep-ph]}}.

\bibitem{Zerwekh:2011wf}
A.~R. Zerwekh, ``{The Axigluon, a Four-Site Model and the Top Quark
  Forward-Backward Asymmetry at the Tevatron},''
  \href{http://dx.doi.org/10.1016/j.physletb.2011.08.064}{{\em Phys.Lett.}
  {\bfseries B704} (2011) 62--65},
\href{http://arxiv.org/abs/1103.0956}{{\ttfamily arXiv:1103.0956 [hep-ph]}}.

\bibitem{Haisch:2011up}
U.~Haisch and S.~Westhoff, ``{Massive Color-Octet Bosons: Bounds on Effects in
  Top-Quark Pair Production},''
  \href{http://dx.doi.org/10.1007/JHEP08(2011)088}{{\em JHEP} {\bfseries 1108}
  (2011) 088},
\href{http://arxiv.org/abs/1106.0529}{{\ttfamily arXiv:1106.0529 [hep-ph]}}.

\bibitem{Tavares:2011zg}
G.~M. Tavares and M.~Schmaltz, ``{Explaining the $t\bar{t}$ asymmetry with a
  light axigluon},'' \href{http://dx.doi.org/10.1103/PhysRevD.84.054008}{{\em
  Phys.Rev.} {\bfseries D84} (2011) 054008},
\href{http://arxiv.org/abs/1107.0978}{{\ttfamily arXiv:1107.0978 [hep-ph]}}.

\bibitem{Alvarez:2011hi}
E.~Alvarez, L.~{Da Rold}, J.~I.~S. Vietto, and A.~Szynkman, ``{Phenomenology of
  a light gluon resonance in top-physics at Tevatron and LHC},''
  \href{http://dx.doi.org/10.1007/JHEP09(2011)007}{{\em JHEP} {\bfseries 1109}
  (2011) 007},
\href{http://arxiv.org/abs/1107.1473}{{\ttfamily arXiv:1107.1473 [hep-ph]}}.

\bibitem{AguilarSaavedra:2011ci}
J.~Aguilar-Saavedra and M.~Perez-Victoria, ``{Shaping the top asymmetry},''
  \href{http://dx.doi.org/10.1016/j.physletb.2011.10.004}{{\em Phys.Lett.}
  {\bfseries B705} (2011) 228--234},
\href{http://arxiv.org/abs/1107.2120}{{\ttfamily arXiv:1107.2120 [hep-ph]}}.

\bibitem{Djouadi:2009nb}
A.~Djouadi, G.~Moreau, F.~Richard, and R.~K. Singh, ``{The Forward-backward
  asymmetry of top quark production at the Tevatron in warped extra dimensional
  models},'' \href{http://dx.doi.org/10.1103/PhysRevD.82.071702}{{\em
  Phys.Rev.} {\bfseries D82} (2010) 071702},
\href{http://arxiv.org/abs/0906.0604}{{\ttfamily arXiv:0906.0604 [hep-ph]}}.

\bibitem{Bauer:2010iq}
M.~Bauer, F.~Goertz, U.~Haisch, T.~Pfoh, and S.~Westhoff, ``{Top-Quark
  Forward-Backward Asymmetry in Randall-Sundrum Models Beyond the Leading
  Order},'' \href{http://dx.doi.org/10.1007/JHEP11(2010)039}{{\em JHEP}
  {\bfseries 1011} (2010) 039},
\href{http://arxiv.org/abs/1008.0742}{{\ttfamily arXiv:1008.0742 [hep-ph]}}.

\bibitem{Delaunay:2011vv}
C.~Delaunay, O.~Gedalia, S.~J. Lee, G.~Perez, and E.~Ponton, ``{Extraordinary
  Phenomenology from Warped Flavor Triviality},''
  \href{http://dx.doi.org/10.1016/j.physletb.2011.08.031}{{\em Phys.Lett.}
  {\bfseries B703} (2011) 486--490},
\href{http://arxiv.org/abs/1101.2902}{{\ttfamily arXiv:1101.2902 [hep-ph]}}.

\bibitem{Park:2009cs}
S.~C. Park and J.~Shu, ``{Split Universal Extra Dimensions and Dark Matter},''
  \href{http://dx.doi.org/10.1103/PhysRevD.79.091702}{{\em Phys.Rev.}
  {\bfseries D79} (2009) 091702},
\href{http://arxiv.org/abs/0901.0720}{{\ttfamily arXiv:0901.0720 [hep-ph]}}.

\bibitem{Chen:2010hm}
C.-H. Chen, G.~Cvetic, and C.~Kim, ``{Forward-backward asymmetry of top quark
  in unparticle physics},''
  \href{http://dx.doi.org/10.1016/j.physletb.2010.10.030}{{\em Phys.Lett.}
  {\bfseries B694} (2011) 393--397},
\href{http://arxiv.org/abs/1009.4165}{{\ttfamily arXiv:1009.4165 [hep-ph]}}.

\bibitem{Alvarez:2010js}
E.~Alvarez, L.~{Da Rold}, and A.~Szynkman, ``{A composite Higgs model analysis
  of forward-backward asymmetries in the production of tops at Tevatron and
  bottoms at LEP and SLC},''
  \href{http://dx.doi.org/10.1007/JHEP05(2011)070}{{\em JHEP} {\bfseries 1105}
  (2011) 070},
\href{http://arxiv.org/abs/1011.6557}{{\ttfamily arXiv:1011.6557 [hep-ph]}}.

\bibitem{Barreto:2011au}
E.~R. Barreto, Y.~Coutinho, and J.~{Sa Borges}, ``{Top quark forward-backward
  asymmetry from the $3-3-1$ model},''
  \href{http://dx.doi.org/10.1103/PhysRevD.83.054006}{{\em Phys.Rev.}
  {\bfseries D83} (2011) 054006},
\href{http://arxiv.org/abs/1103.1266}{{\ttfamily arXiv:1103.1266 [hep-ph]}}.

\bibitem{Foot:2011xu}
R.~Foot, ``{Top quark forward-backward asymmetry from SU($N_c$) color},''
  \href{http://dx.doi.org/10.1103/PhysRevD.83.114013}{{\em Phys.Rev.}
  {\bfseries D83} (2011) 114013},
\href{http://arxiv.org/abs/1103.1940}{{\ttfamily arXiv:1103.1940 [hep-ph]}}.

\bibitem{Kamenik:2011wt}
J.~F. Kamenik, J.~Shu, and J.~Zupan, ``{Review of new physics effects in
  $t\bar{t}$ production},''
\href{http://arxiv.org/abs/1107.5257}{{\ttfamily arXiv:1107.5257 [hep-ph]}}.

\bibitem{Ligeti:2011vt}
Z.~Ligeti, G.~M. Tavares, and M.~Schmaltz, ``{Explaining the $t \bar{t}$
  forward-backward asymmetry without dijet or flavor anomalies},''
  \href{http://dx.doi.org/10.1007/JHEP06(2011)109}{{\em JHEP} {\bfseries 1106}
  (2011) 109},
\href{http://arxiv.org/abs/1103.2757}{{\ttfamily arXiv:1103.2757 [hep-ph]}}.

\bibitem{Grinstein:2011yv}
B.~Grinstein, A.~L. Kagan, M.~Trott, and J.~Zupan, ``{Forward-backward
  asymmetry in $t \bar{t}$ production from flavour symmetries},''
  \href{http://dx.doi.org/10.1103/PhysRevLett.107.012002}{{\em Phys.Rev.Lett.}
  {\bfseries 107} (2011) 012002},
\href{http://arxiv.org/abs/1102.3374}{{\ttfamily arXiv:1102.3374 [hep-ph]}}.

\bibitem{Nelson:2011us}
A.~E. Nelson, T.~Okui, and T.~S. Roy, ``{A unified, flavor symmetric
  explanation for the $t \bar{t}$ asymmetry and Wjj excess at CDF},''
  \href{http://dx.doi.org/10.1103/PhysRevD.84.094007}{{\em Phys.Rev.}
  {\bfseries D84} (2011) 094007},
  \href{http://arxiv.org/abs/1104.2030}{{\ttfamily arXiv:1104.2030 [hep-ph]}}.
minor changes to text, references added, version to be published in PRD.

\bibitem{Bauer:2009cc}
C.~W. Bauer, Z.~Ligeti, M.~Schmaltz, J.~Thaler, and D.~G. Walker,
  ``{Supermodels for early LHC},''
  \href{http://dx.doi.org/10.1016/j.physletb.2010.05.032}{{\em Phys.Lett.}
  {\bfseries B690} (2010) 280--288},
\href{http://arxiv.org/abs/0909.5213}{{\ttfamily arXiv:0909.5213 [hep-ph]}}.

\bibitem{Arnold:2009ay}
J.~M. Arnold, M.~Pospelov, M.~Trott, and M.~B. Wise, ``{Scalar Representations
  and Minimal Flavor Violation},''
  \href{http://dx.doi.org/10.1007/JHEP01(2010)073}{{\em JHEP} {\bfseries 1001}
  (2010) 073},
\href{http://arxiv.org/abs/0911.2225}{{\ttfamily arXiv:0911.2225 [hep-ph]}}.

\bibitem{Grinstein:2011dz}
B.~Grinstein, A.~L. Kagan, J.~Zupan, and M.~Trott, ``{Flavor Symmetric Sectors
  and Collider Physics},''
  \href{http://dx.doi.org/10.1007/JHEP10(2011)072}{{\em JHEP} {\bfseries 1110}
  (2011) 072},
\href{http://arxiv.org/abs/1108.4027}{{\ttfamily arXiv:1108.4027 [hep-ph]}}.

\bibitem{Craig:2011an}
N.~Craig, C.~Kilic, and M.~J. Strassler, ``{LHC Charge Asymmetry as Constraint
  on Models for the Tevatron Top Anomaly},''
  \href{http://dx.doi.org/10.1103/PhysRevD.84.035012}{{\em Phys.Rev.}
  {\bfseries D84} (2011) 035012},
\href{http://arxiv.org/abs/1103.2127}{{\ttfamily arXiv:1103.2127 [hep-ph]}}.

\bibitem{Bhattacherjee:2011nr}
B.~Bhattacherjee, S.~S. Biswal, and D.~Ghosh, ``{Top quark forward-backward
  asymmetry at Tevatron and its implications at the LHC},''
  \href{http://dx.doi.org/10.1103/PhysRevD.83.091501}{{\em Phys.Rev.}
  {\bfseries D83} (2011) 091501},
\href{http://arxiv.org/abs/1102.0545}{{\ttfamily arXiv:1102.0545 [hep-ph]}}.

\bibitem{Jung:2009jz}
S.~Jung, H.~Murayama, A.~Pierce, and J.~D. Wells, ``{Top quark forward-backward
  asymmetry from new t-channel physics},''
  \href{http://dx.doi.org/10.1103/PhysRevD.81.015004}{{\em Phys.Rev.}
  {\bfseries D81} (2010) 015004},
\href{http://arxiv.org/abs/0907.4112}{{\ttfamily arXiv:0907.4112 [hep-ph]}}.

\bibitem{Ko:2011vd}
P.~Ko, Y.~Omura, and C.~Yu, ``{Top forward-backward asymmetry and the CDF Wjj
  excess in leptophobic U(1)' flavor models},''
\href{http://arxiv.org/abs/1108.0350}{{\ttfamily arXiv:1108.0350 [hep-ph]}}.

\bibitem{Cheung:2009ch}
K.~Cheung, W.-Y. Keung, and T.-C. Yuan, ``{Top Quark Forward-Backward
  Asymmetry},'' \href{http://dx.doi.org/10.1016/j.physletb.2009.11.015}{{\em
  Phys.Lett.} {\bfseries B682} (2009) 287--290},
\href{http://arxiv.org/abs/0908.2589}{{\ttfamily arXiv:0908.2589 [hep-ph]}}.

\bibitem{Xiao:2010hm}
B.~Xiao, Y.~kai Wang, and S.~hua Zhu, ``{Forward-backward Asymmetry and
  Differential Cross Section of Top Quark in Flavor Violating Z' model at
  $\mathcal{O}(\alpha_{s}^{2} \alpha_{X})$},''
  \href{http://dx.doi.org/10.1103/PhysRevD.82.034026}{{\em Phys.Rev.}
  {\bfseries D82} (2010) 034026},
\href{http://arxiv.org/abs/1006.2510}{{\ttfamily arXiv:1006.2510 [hep-ph]}}.

\bibitem{Cheung:2011qa}
K.~Cheung and T.-C. Yuan, ``{Top Quark Forward-Backward Asymmetry in the Large
  Invariant Mass Region},''
  \href{http://dx.doi.org/10.1103/PhysRevD.83.074006}{{\em Phys.Rev.}
  {\bfseries D83} (2011) 074006},
\href{http://arxiv.org/abs/1101.1445}{{\ttfamily arXiv:1101.1445 [hep-ph]}}.

\bibitem{Berger:2011ua}
E.~L. Berger, Q.-H. Cao, C.-R. Chen, C.~S. Li, and H.~Zhang, ``{Top Quark
  Forward-Backward Asymmetry and Same-Sign Top Quark Pairs},''
  \href{http://dx.doi.org/10.1103/PhysRevLett.106.201801}{{\em Phys.Rev.Lett.}
  {\bfseries 106} (2011) 201801},
\href{http://arxiv.org/abs/1101.5625}{{\ttfamily arXiv:1101.5625 [hep-ph]}}.

\bibitem{Duraisamy:2011pt}
M.~Duraisamy, A.~Rashed, and A.~Datta, ``{The Top Forward Backward Asymmetry
  with general Z$^{\prime}$ couplings},''
  \href{http://dx.doi.org/10.1103/PhysRevD.84.054018}{{\em Phys.Rev.}
  {\bfseries D84} (2011) 054018},
  \href{http://arxiv.org/abs/1106.5982}{{\ttfamily arXiv:1106.5982 [hep-ph]}}.
13 pages, 14 figures, minor typos corrected, accepted for publication in
  Physical Review D.

\bibitem{Cao:2011ew}
J.~Cao, L.~Wang, L.~Wu, and J.~M. Yang, ``{Top quark forward-backward
  asymmetry, FCNC decays and like-sign pair production as a joint probe of new
  physics},'' \href{http://dx.doi.org/10.1103/PhysRevD.84.074001}{{\em
  Phys.Rev.} {\bfseries D84} (2011) 074001},
\href{http://arxiv.org/abs/1101.4456}{{\ttfamily arXiv:1101.4456 [hep-ph]}}.

\bibitem{Chen:2011mga}
C.-H. Chen, S.~S. Law, and R.-H. Li, ``{Rare B decays and Tevatron top-pair
  asymmetry},'' \href{http://dx.doi.org/10.1088/0954-3899/38/11/115008}{{\em
  J.Phys.G} {\bfseries G38} (2011) 115008},
\href{http://arxiv.org/abs/1104.1497}{{\ttfamily arXiv:1104.1497 [hep-ph]}}.

\bibitem{Barger:2011ih}
V.~Barger, W.-Y. Keung, and C.-T. Yu, ``{Tevatron Asymmetry of Tops in a
  W$^{\prime}$, Z$^{\prime}$ Model},''
  \href{http://dx.doi.org/10.1016/j.physletb.2011.03.010}{{\em Phys.Lett.}
  {\bfseries B698} (2011) 243--250},
\href{http://arxiv.org/abs/1102.0279}{{\ttfamily arXiv:1102.0279 [hep-ph]}}.

\bibitem{Frank:2011rb}
M.~Frank, A.~Hayreter, and I.~Turan, ``{Top Quark Pair Production and Asymmetry
  at the Tevatron and LHC in Left-Right Models},''
  \href{http://dx.doi.org/10.1103/PhysRevD.84.114007}{{\em Phys.Rev.}
  {\bfseries D84} (2011) 114007},
\href{http://arxiv.org/abs/1108.0998}{{\ttfamily arXiv:1108.0998 [hep-ph]}}.

\bibitem{Barger:2010mw}
V.~Barger, W.-Y. Keung, and C.-T. Yu, ``{Asymmetric Left-Right Model and the
  Top Pair Forward-Backward Asymmetry},''
  \href{http://dx.doi.org/10.1103/PhysRevD.81.113009}{{\em Phys.Rev.}
  {\bfseries D81} (2010) 113009},
\href{http://arxiv.org/abs/1002.1048}{{\ttfamily arXiv:1002.1048 [hep-ph]}}.

\bibitem{Shelton:2011hq}
J.~Shelton and K.~M. Zurek, ``{Maximal flavor violation from new right-handed
  gauge bosons},'' \href{http://dx.doi.org/10.1103/PhysRevD.83.091701}{{\em
  Phys.Rev.} {\bfseries D83} (2011) 091701},
\href{http://arxiv.org/abs/1101.5392}{{\ttfamily arXiv:1101.5392 [hep-ph]}}.

\bibitem{Aaltonen:2009iz}
{\bfseries CDF Collaboration} Collaboration, T.~Aaltonen {\em et~al.}, ``{First
  Measurement of the $t \bar{t}$ Differential Cross Section
  $\frac{d\sigma}{dM_{t\bar{t}}}$ in $p \bar{t}$ Collisions at $\sqrt{s}=$1.96
  TeV},'' \href{http://dx.doi.org/10.1103/PhysRevLett.102.222003}{{\em
  Phys.Rev.Lett.} {\bfseries 102} (2009) 222003},
\href{http://arxiv.org/abs/0903.2850}{{\ttfamily arXiv:0903.2850 [hep-ex]}}.

\bibitem{Ahrens:2011mw}
V.~Ahrens, A.~Ferroglia, M.~Neubert, B.~D. Pecjak, and L.-L. Yang,
  ``{RG-improved single-particle inclusive cross sections and forward-backward
  asymmetry in $t\bar t$ production at hadron colliders},''
  \href{http://dx.doi.org/10.1007/JHEP09(2011)070}{{\em JHEP} {\bfseries 1109}
  (2011) 070},
\href{http://arxiv.org/abs/1103.0550}{{\ttfamily arXiv:1103.0550 [hep-ph]}}.

\bibitem{Martin:2009iq}
A.~Martin, W.~Stirling, R.~Thorne, and G.~Watt, ``{Parton distributions for the
  LHC},'' \href{http://dx.doi.org/10.1140/epjc/s10052-009-1072-5}{{\em
  Eur.Phys.J.} {\bfseries C63} (2009) 189--285},
\href{http://arxiv.org/abs/0901.0002}{{\ttfamily arXiv:0901.0002 [hep-ph]}}.

\bibitem{Gresham:2011pa}
M.~I. Gresham, I.-W. Kim, and K.~M. Zurek, ``{On Models of New Physics for the
  Tevatron Top $A_{FB}$},''
  \href{http://dx.doi.org/10.1103/PhysRevD.83.114027}{{\em Phys.Rev.}
  {\bfseries D83} (2011) 114027},
\href{http://arxiv.org/abs/1103.3501}{{\ttfamily arXiv:1103.3501 [hep-ph]}}.

\bibitem{Blum:2011fa}
K.~Blum, Y.~Hochberg, and Y.~Nir, ``{Scalar-mediated $t\bar t$ forward-backward
  asymmetry},'' \href{http://dx.doi.org/10.1007/JHEP10(2011)124}{{\em JHEP}
  {\bfseries 1110} (2011) 124},
  \href{http://arxiv.org/abs/1107.4350}{{\ttfamily arXiv:1107.4350 [hep-ph]}}.
22 pages, 1 figure and 2 tables. v2: Corrected Eqs.(50,51,74), adapted Fig.1,
  Tab.1 and relevant discussions. Extended discussion of top decay and single
  top.

\bibitem{ATLAS:2012an}
{\bfseries ATLAS Collaboration} Collaboration, G.~Aad {\em et~al.},
  ``{Measurement of the charge asymmetry in top quark pair production in pp
  collisions at sqrt(s) = 7 TeV using the ATLAS detector},''
  \href{http://dx.doi.org/10.1140/epjc/s10052-012-2039-5}{{\em Eur.Phys.J.}
  {\bfseries C72} (2012) 2039},
\href{http://arxiv.org/abs/1203.4211}{{\ttfamily arXiv:1203.4211 [hep-ex]}}.

\bibitem{Rodrigo:2012as}
G.~Rodrigo, ``{The ttbar asymmetry in the Standard Model and beyond},''
\href{http://arxiv.org/abs/1207.0331}{{\ttfamily arXiv:1207.0331 [hep-ph]}}.

\bibitem{Allanach:2012tc}
B.~Allanach and Sridhar, ``{A Supersymmetric Explanation for High $A_{FB}(t\bar
  t)$ Via R-Parity Violation},''
\href{http://arxiv.org/abs/1205.5170}{{\ttfamily arXiv:1205.5170 [hep-ph]}}.

\bibitem{Hagiwara:2012gy}
K.~Hagiwara and J.~Nakamura, ``{Diquark contributions to Top quark charge
  asymmetry at the Tevatron and LHC},''
\href{http://arxiv.org/abs/1205.5005}{{\ttfamily arXiv:1205.5005 [hep-ph]}}.

\bibitem{Chang:1996sw}
D.~Chang and W.-Y. Keung, ``{New limits on R-parity breakings in supersymmetric
  standard models},''
  \href{http://dx.doi.org/10.1016/S0370-2693(96)01271-3}{{\em Phys.Lett.}
  {\bfseries B389} (1996) 294--298},
\href{http://arxiv.org/abs/hep-ph/9608313}{{\ttfamily arXiv:hep-ph/9608313
  [hep-ph]}}.

\bibitem{Cline:1990bw}
J.~M. Cline and S.~Raby, ``{Gravitino induced baryogenesis: A Problem made a
  virtue},''
\href{http://dx.doi.org/10.1103/PhysRevD.43.1781}{{\em Phys.Rev.} {\bfseries
  D43} (1991) 1781--1787}.

\bibitem{Gresham:2012wc}
M.~I. Gresham, I.-W. Kim, S.~Tulin, and K.~M. Zurek, ``{Confronting Top AFB
  with Parity Violation Constraints},''
  \href{http://arxiv.org/abs/1203.1320}{{\ttfamily arXiv:1203.1320 [hep-ph]}}.
4 pages.

\bibitem{Arhrib:2009hu}
A.~Arhrib, R.~Benbrik, and C.-H. Chen, ``{Forward-backward asymmetry of top
  quark in diquark models},''
  \href{http://dx.doi.org/10.1103/PhysRevD.82.034034}{{\em Phys.Rev.}
  {\bfseries D82} (2010) 034034},
\href{http://arxiv.org/abs/0911.4875}{{\ttfamily arXiv:0911.4875 [hep-ph]}}.

\bibitem{Manohar:2006ga}
A.~V. Manohar and M.~B. Wise, ``{Flavor changing neutral currents, an extended
  scalar sector, and the Higgs production rate at the CERN LHC},''
  \href{http://dx.doi.org/10.1103/PhysRevD.74.035009}{{\em Phys.Rev.}
  {\bfseries D74} (2006) 035009},
\href{http://arxiv.org/abs/hep-ph/0606172}{{\ttfamily arXiv:hep-ph/0606172
  [hep-ph]}}.

\bibitem{Bouchiat:1997mj}
M.~Bouchiat and C.~Bouchiat, ``{Parity violation in atoms},''
\href{http://dx.doi.org/10.1088/0034-4885/60/11/004}{{\em Rept.Prog.Phys.}
  {\bfseries 60} (1997) 1351--1396}.

\bibitem{PDG}
K.~Nakamura {\em et~al.} {\em J. Phys G.} {\bfseries G 37} (2010) 075021.

\bibitem{Porsev:2009pr}
S.~Porsev, K.~Beloy, and A.~Derevianko, ``{Precision determination of
  electroweak coupling from atomic parity violation and implications for
  particle physics},''
  \href{http://dx.doi.org/10.1103/PhysRevLett.102.181601}{{\em Phys.Rev.Lett.}
  {\bfseries 102} (2009) 181601},
\href{http://arxiv.org/abs/0902.0335}{{\ttfamily arXiv:0902.0335 [hep-ph]}}.

\bibitem{Porsev:2010de}
S.~Porsev, K.~Beloy, and A.~Derevianko, ``{Precision determination of weak
  charge of $^{133}$Cs from atomic parity violation},''
  \href{http://dx.doi.org/10.1103/PhysRevD.82.036008}{{\em Phys.Rev.}
  {\bfseries D82} (2010) 036008},
\href{http://arxiv.org/abs/1006.4193}{{\ttfamily arXiv:1006.4193 [hep-ph]}}.

\bibitem{Carlini:2010zz}
R.~Carlini, ``{The Qweak experiment: A precision measurement of the proton's
  weak charge},''
{\em AIP Conf.Proc.} {\bfseries 1261} (2010) 172--178.

\bibitem{Young:2007zs}
R.~D. Young, R.~D. Carlini, A.~W. Thomas, and J.~Roche, ``{Testing the standard
  model by precision measurement of the weak charges of quarks},''
  \href{http://dx.doi.org/10.1103/PhysRevLett.99.122003}{{\em Phys.Rev.Lett.}
  {\bfseries 99} (2007) 122003},
\href{http://arxiv.org/abs/0704.2618}{{\ttfamily arXiv:0704.2618 [hep-ph]}}.

\bibitem{Aaltonen:2011em}
{\bfseries CDF Collaboration} Collaboration, T.~Aaltonen {\em et~al.},
  ``{Measurement of the Top Quark Mass in the All-Hadronic Mode at CDF},''
\href{http://arxiv.org/abs/1112.4891}{{\ttfamily arXiv:1112.4891 [hep-ex]}}.

\bibitem{ATLAS:2012hg}
{\bfseries ATLAS Collaboration} Collaboration, G.~Aad {\em et~al.},
  ``{Measurements of top quark pair relative differential cross-sections with
  ATLAS in pp collisions at sqrt(s) = 7 TeV},''
\href{http://arxiv.org/abs/1207.5644}{{\ttfamily arXiv:1207.5644 [hep-ex]}}.

\bibitem{Manohar:2012rs}
A.~V. Manohar and M.~Trott, ``{Electroweak Sudakov Corrections and the Top
  Quark Forward-Backward Asymmetry},''
  \href{http://arxiv.org/abs/1201.3926}{{\ttfamily arXiv:1201.3926 [hep-ph]}}.
5 pages, 2 figures.

\bibitem{AguilarSaavedra:2011vw}
J.~Aguilar-Saavedra and M.~Perez-Victoria, ``{Probing the Tevatron t tbar
  asymmetry at LHC},'' \href{http://dx.doi.org/10.1007/JHEP05(2011)034}{{\em
  JHEP} {\bfseries 1105} (2011) 034},
\href{http://arxiv.org/abs/1103.2765}{{\ttfamily arXiv:1103.2765 [hep-ph]}}.

\bibitem{Bona:2007vi}
{\bfseries UTfit Collaboration} Collaboration, M.~Bona {\em et~al.},
  ``{Model-independent constraints on $\Delta$ F=2 operators and the scale of
  new physics},'' \href{http://dx.doi.org/10.1088/1126-6708/2008/03/049}{{\em
  JHEP} {\bfseries 0803} (2008) 049},
\href{http://arxiv.org/abs/0707.0636}{{\ttfamily arXiv:0707.0636 [hep-ph]}}.

\bibitem{Dimopoulos:1995mi}
S.~Dimopoulos and G.~Giudice, ``{Naturalness constraints in supersymmetric
  theories with nonuniversal soft terms},''
  \href{http://dx.doi.org/10.1016/0370-2693(95)00961-J}{{\em Phys.Lett.}
  {\bfseries B357} (1995) 573--578},
\href{http://arxiv.org/abs/hep-ph/9507282}{{\ttfamily arXiv:hep-ph/9507282
  [hep-ph]}}.

\bibitem{Cohen:1996vb}
A.~G. Cohen, D.~Kaplan, and A.~Nelson, ``{The More minimal supersymmetric
  standard model},''
  \href{http://dx.doi.org/10.1016/S0370-2693(96)01183-5}{{\em Phys.Lett.}
  {\bfseries B388} (1996) 588--598},
\href{http://arxiv.org/abs/hep-ph/9607394}{{\ttfamily arXiv:hep-ph/9607394
  [hep-ph]}}.

\bibitem{Barbieri:2010pd}
R.~Barbieri, E.~Bertuzzo, M.~Farina, P.~Lodone, and D.~Pappadopulo, ``{A Non
  Standard Supersymmetric Spectrum},''
  \href{http://dx.doi.org/10.1007/JHEP08(2010)024}{{\em JHEP} {\bfseries 1008}
  (2010) 024},
\href{http://arxiv.org/abs/1004.2256}{{\ttfamily arXiv:1004.2256 [hep-ph]}}.

\bibitem{Craig:2011yk}
N.~Craig, D.~Green, and A.~Katz, ``{(De)Constructing a Natural and Flavorful
  Supersymmetric Standard Model},''
  \href{http://dx.doi.org/10.1007/JHEP07(2011)045}{{\em JHEP} {\bfseries 1107}
  (2011) 045},
\href{http://arxiv.org/abs/1103.3708}{{\ttfamily arXiv:1103.3708 [hep-ph]}}.

\bibitem{Brust:2012uf}
C.~Brust, A.~Katz, and R.~Sundrum, ``{SUSY Stops at a Bump},''
  \href{http://dx.doi.org/10.1007/JHEP08(2012)059}{{\em JHEP} {\bfseries 1208}
  (2012) 059},
\href{http://arxiv.org/abs/1206.2353}{{\ttfamily arXiv:1206.2353 [hep-ph]}}.

\bibitem{Brust:2011tb}
C.~Brust, A.~Katz, S.~Lawrence, and R.~Sundrum, ``{SUSY, the Third Generation
  and the LHC},'' \href{http://dx.doi.org/10.1007/JHEP03(2012)103}{{\em JHEP}
  {\bfseries 1203} (2012) 103},
\href{http://arxiv.org/abs/1110.6670}{{\ttfamily arXiv:1110.6670 [hep-ph]}}.

\bibitem{Cao:2009uz}
J.~Cao, Z.~Heng, L.~Wu, and J.~M. Yang, ``{Top quark forward-backward asymmetry
  at the Tevatron: A Comparative study in different new physics models},''
  \href{http://dx.doi.org/10.1103/PhysRevD.81.014016}{{\em Phys.Rev.}
  {\bfseries D81} (2010) 014016},
\href{http://arxiv.org/abs/0912.1447}{{\ttfamily arXiv:0912.1447 [hep-ph]}}.

\bibitem{Goity:1994dq}
J.~Goity and M.~Sher, ``{Bounds on delta B = 1 couplings in the supersymmetric
  standard model},'' \href{http://dx.doi.org/10.1016/0370-2693(94)01688-9}{{\em
  Phys.Lett.} {\bfseries B346} (1995) 69--74},
\href{http://arxiv.org/abs/hep-ph/9412208}{{\ttfamily arXiv:hep-ph/9412208
  [hep-ph]}}.

\bibitem{Lane:1991qh}
K.~D. Lane and M.~Ramana, ``{Walking technicolor signatures at hadron
  colliders},''
\href{http://dx.doi.org/10.1103/PhysRevD.44.2678}{{\em Phys.Rev.} {\bfseries
  D44} (1991) 2678--2700}.

\end{thebibliography}\endgroup

\end{document}